\PassOptionsToPackage{prologue, dvipsnames}{xcolor}
\documentclass[sigconf]{acmart}
\usepackage{soul}
\usepackage{float}
\usepackage{stfloats}
\usepackage{tablefootnote}
\usepackage{mathtools}
\usepackage{amsfonts}
\usepackage{multirow}
\usepackage{multicol}
\usepackage{booktabs}
\usepackage{listings}
\usepackage{tabularray}
\usepackage{algorithm} % For algorithm
\usepackage{algpseudocode}
\usepackage{cleveref}
\usepackage{makecell}
\usepackage{tablefootnote}
\usepackage{color, colortbl}
\definecolor{Gray}{gray}{0.9}
\definecolor{darkpastelgreen}{rgb}{0.01, 0.75, 0.24}
\definecolor{cadetgrey}{rgb}{0.57, 0.64, 0.69}
\definecolor{camel}{rgb}{0.76, 0.6, 0.42}
\definecolor{lightskyblue}{rgb}{0.53, 0.81, 0.98}
\definecolor{lightsblue}{rgb}{0.53, 0.81, 0.98}
\definecolor{lightblue}{rgb}{0.68, 0.85, 0.9}
\definecolor{softblue}{rgb}{0.85, 0.91, 0.98}
\definecolor{mygray}{gray}{0.6}
\usepackage{mathtools} % for \mathrlap
\usepackage[english]{babel}
\usepackage{amsthm}
\usepackage{etoolbox}
\usepackage{svg}
\usepackage{enumitem}

\newtheorem*{remark}{Remark}

\AtBeginDocument{%
  }

\setcopyright{acmlicensed}
\copyrightyear{2025}
\acmYear{2025}
\acmDOI{XXXXXXX.XXXXXXX}

\acmConference[KDD '26]{32nd ACM SIGKDD Conference on Knowledge Discovery and Data Mining}{August 9--13,
  2026}{Jeju, Korea}
\acmISBN{978-1-4503-XXXX-X/2018/06}

\makeatletter
\gdef\@copyrightpermission{
  \begin{minipage}{0.3\columnwidth}
   \href{https://creativecommons.org/licenses/by/4.0/}{\includegraphics[width=0.90\textwidth]{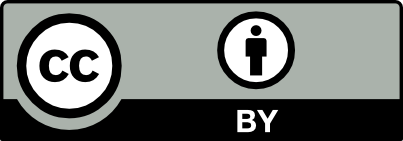}}
  \end{minipage}\hfill
  \begin{minipage}{0.7\columnwidth}
   \href{https://creativecommons.org/licenses/by/4.0/}{This work is licensed under a Creative Commons Attribution International 4.0 License.}
  \end{minipage}
  \vspace{5pt}
}
\makeatother
\begin{document}
\setlength\abovedisplayskip{3pt}
\setlength\belowdisplayskip{3pt}
\title{Towards A Tri-View Diffusion Framework for Recommendation}

\author{Ximing Chen}
\affiliation{%
  \institution{University of Macau}
  \country{Macau SAR, China}
}
\email{yc37921@um.edu.mo}

\author{Pui Ieng Lei}
\affiliation{%
  \institution{University of Macau}
  \city{Macau SAR}
  \country{China}
}
\email{yc37460@um.edu.mo}

\author{Yijun Sheng}
\affiliation{%
  \institution{University of Macau}
  \city{Macau SAR}
  \country{China}
}
\email{yc17419@um.edu.mo}

\author{Yanyan Liu}
\affiliation{%
  \institution{University of Macau}
  \city{Macau SAR}
  \country{China}
}
\email{yc07402@um.edu.mo}

\author{Zhiguo Gong}
\authornote{Corresponding author.}
\affiliation{%
  \institution{University of Macau}
  \city{Macau SAR}
  \country{China}
}
\email{fstzgg@um.edu.mo}

% \affiliation{%

% \affiliation{%

% \affiliation{%

% \affiliation{%

% \affiliation{%

% \affiliation{%

% \affiliation{%

\renewcommand{\shortauthors}{Ximing Chen, Pui Ieng Lei, Yijun Sheng, Yanyan Liu, \& Zhiguo Gong}

\begin{abstract}
Diffusion models (DMs) have recently gained significant interest for their exceptional potential in recommendation tasks. This stems primarily from their prominent capability in distilling, modeling, and generating comprehensive user preferences. However, previous work fails to examine DMs in recommendation tasks through a rigorous lens. In this paper, we first experimentally investigate the completeness of recommender models from a thermodynamic view. We reveal that existing DM-based recommender models operate by maximizing the \textit{energy}, while classic recommender models operate by reducing the \textit{entropy}. Based on this finding, we propose a minimalistic diffusion framework that incorporates both factors via the maximization of \textit{Helmholtz free energy}. Meanwhile, to foster the optimization, our reverse process is armed with a well-designed denoiser to maintain the inherent anisotropy, which measures the user-item cross-correlation in the context of bipartite graphs. Finally, we adopt an Acceptance-Rejection Gumbel Sampling Process (AR-GSP) to prioritize the far-outnumbered unobserved interactions for model robustness. AR-GSP integrates an acceptance-rejection sampling to ensure high-quality hard negative samples for general recommendation tasks, and a timestep-dependent Gumbel Softmax to handle an adaptive sampling strategy for diffusion models. Theoretical analyses and extensive experiments demonstrate that our proposed framework has distinct superiority over baselines in terms of accuracy and efficiency. 
\end{abstract}




\maketitle

% \vspace{-0.5\baselineskip}
\section{Introduction}
Recommendation systems persist as an escalating and influential research field in response to information overload, yet confront mounting challenges, such as data sparsity and diversity. Over these years, diffusion models (DMs)~\cite{DDPM,SDE,MultiDiff} have established themselves as powerful tools in recommendations tasks~\cite{CODIGEM,DiffRec,BSPM}, demonstrating remarkable capability in noise smoothing, dimensionality reduction, and prediction generation. Various diffusion-based recommender approaches~\cite{Survey_DiffRec} - including social recommendation~\cite{GDSSL,RecDiff}, sequential recommendation~\cite{DreamRec,PreferDiff}, and cross-domain recommendation~\cite{CausalDiffRec,DMCDR} - have achieved notable prominence. Despite their success, existing diffusion-based recommenders have not rigorously examined DMs in recommendation tasks.

\begin{figure}[!t]
  \centering
  \vspace{-2mm}
  \includegraphics[width=1.\columnwidth]{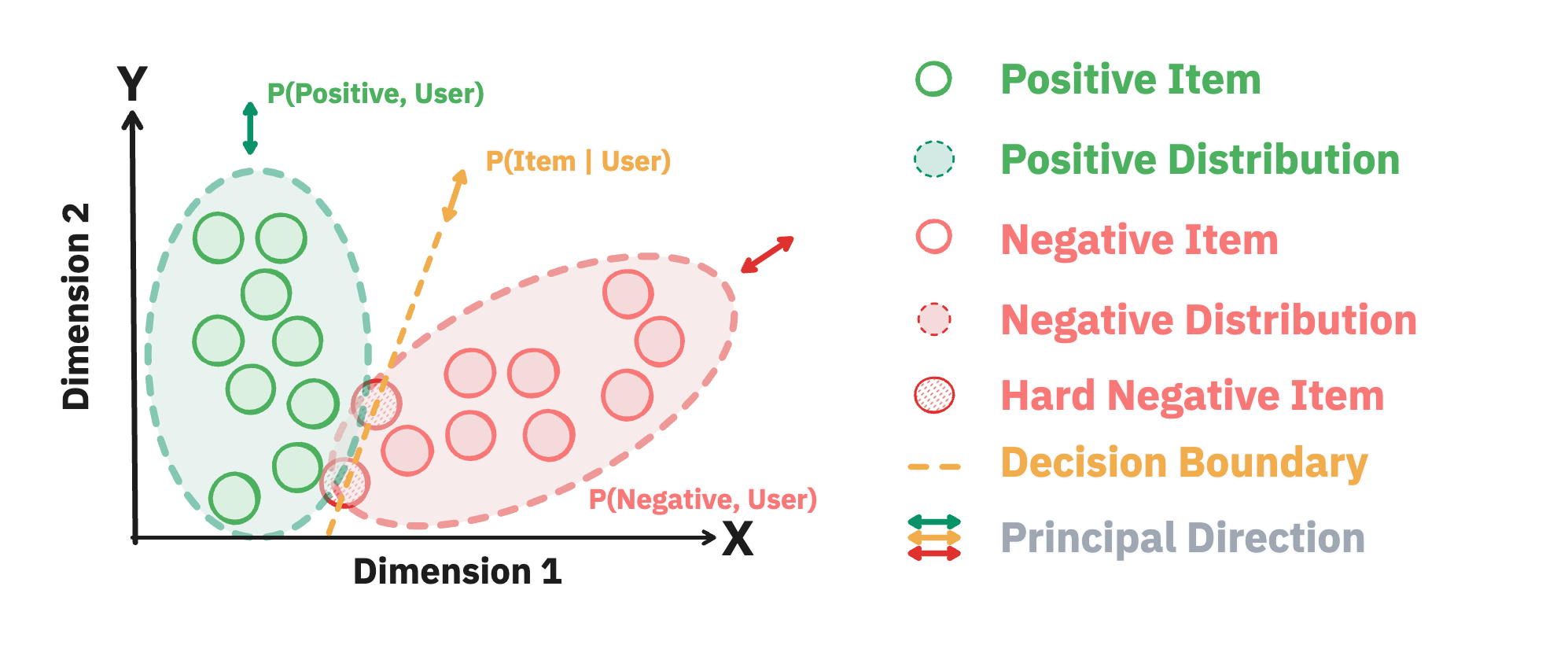}
  \vspace{-8mm}
  % \caption{Toy example of the tri-view of our proposed framework: (a) Thermodynamic View, (b) Topological View, and (c) Hard-Negative View}
  \caption{Toy illustration of 2D visualization on items.}
  \label{fig:motivation}
  \vspace{-6mm}
\end{figure}

A primary focus of existing DM-based recommender models is their reliance on the denoising and regenerating mechanism in the reverse process. Intuitively, researchers presume that iteratively refining noisy user-item interactions through a gradual denoising process can capture nuanced user preferences more effectively than deterministic and single-step generative approaches, in spite of any loss used for training convergence~\cite{CF_Diff}. However, the latest work~\cite{PreferDiff} has argued that existing DM-based recommender models blindly adopt original objectives (e.g., Mean Squared Error (MSE)) from other fields~\cite{thermoDiff,DDPM}, without examining them from the perspective of recommendation tasks. As shown in Figure~\ref{fig:motivation}, classic recommendation objectives (e.g., Bayesian Personalized Ranking (BPR)~\cite{BPR}) consider the essential role of negative items, which enhances personalization performance by maximizing the decision boundary $P(Item|User)$. On the other hand, MSE is intrinsically capable of mitigating data sparsity with the joint distribution $P(Item, User)$~\cite{MultiVAE}, whereas its potential in ranking implicit feedback is suboptimal~\cite{Wide&Deep}. Although a surrogate optimization target is proposed for sequential recommendations based on BPR and MSE, its elusive design leads to \textit{biphasic oscillation}~\cite{FF} due to the intertwined learnable parameters in the denoiser during training, and fails to converge in general recommendation tasks eventually. Hence, it naturally gives rise to the question: \textit{How to elucidate and utilize the characteristic of losses with concision and profundity?}

To investigate this issue, we first conduct pilot experiments on two genres of models: DiffRec~\cite{DiffRec} for DM-based recommender models, while BPR~\cite{BPR} and LightGCN~\cite{LightGCN} for the classic ones. Two metrics are of consideration: \textit{energy}~\cite{energy} demonstrates the number of reconstructed interactions after training, while \textit{entropy}~\cite{entropy} qualifies the uncertainty of reconstructed items' distribution. The visualization shows that during optimization, DiffRec maximizes an \textit{energy} function, while BPR and LightGCN minimize \textit{entropy}. Therefore, from the thermodynamic view, follow-up questions come up: \textit{(1) How to judiciously design training objectives for two genres in general? and (2) How to organize them effectively and efficiently?}

For the first question, a minimalist diffusion framework that incorporates both genres of objectives via \textit{Helmholtz free energy}~\cite{legendre} is spontaneously proposed. It follows the concept from thermodynamics based on our observations from pilot experiments. However, the limitation of DM-based models in capturing high-order collaborative signals~\cite{BSPM,GiffCF,HDRM} hinders an effective integration of both objectives for the second question. Specifically, the inherent anisotropy of user-item interactions, which manifests the topological information (i.e., principal direction~\cite{PCA} in Figure~\ref{fig:motivation}) embedded in the hyperbolic space~\cite{HGCF}, is impaired by adding progressive isotropic Gaussian noise. Thus, the results might underperform either genre of models as the undertrained decision boundary deteriorates the joint-training of both genres of objectives mutually. 

The demand for robustness of the mutual joint-training objective also necessitates an efficient yet effective negative sampling method. This approach must adequately refine the decision boundary along with the item distribution. However, classic recommender models typically rely on random negative sampling (RNS)~\cite{BPR}, which often yields uninformative negatives. Hard negative sampling (Hard NS)~\cite{MCNS,DMNS} offers an improvement by focusing on samples near the decision boundary (cf. Figure~\ref{fig:motivation}). Yet, it compromises models' ability amid optimization and generalization via a sub-linear correlation with the positive distribution. More critically, the varying magnitude of noise across different timesteps complicates a productive sampling process for diffusion models~\cite{gumbel_softmax}.

To address these issues, we propose \textbf{TV-Diff}, a novel diffusion recommendation framework from a rigorous tri-view lens. TV-Diff employs three mutually interacting components: \textbf{Helmholtz Free Energy Maximization} incorporates designated \textit{energy-based} and \textit{entropy-based} training objectives to optimize the model from a thermodynamic view as a foundation of our proposed framework. \textbf{Anisotropic Denoiser} maintains the anisotropic signals by disentangling the monolithic user-wise encoder-decoder paradigm into users and items aspects, and reconstructing interactions by user-item cross-correlation measurements with explicit topological information. \textbf{Acceptance-Rejection Gumbel Sampling Process} first discards true negatives on the general aspect via an acceptance probability. Then, a timestep-dependent Gumbel Softmax is introduced to buffer the risk of mistaking true positives due to excessive noise in the diffusion process. 
% Generally, our research methodology follows a progressive design paradigm. TV-Diff's foundation lies in the maximization of Helmholtz free energy (View I), which integrates energy- and entropy-based objectives for comprehensive optimization. The anisotropic denoiser (View II) subsequently operationalizes entropy reduction within diffusion architectures. AR-GSP (View III) is also built upon entropy-based objectives, generating sample-wise confidence $c_{u,i}$ through reconstructed interactions from View II.
Rationale analyses of the last two components are also provided in the context of \textit{Helmholtz free energy}. To summarize, key contributions of TV-Diff are:

\setlist[itemize]{leftmargin=3mm}
\begin{itemize}
    % \vspace{-2mm}
    \item We present TV-Diff, a minimalistic diffusion recommender framework, to revisit diffusion models from a thermodynamic view, a topological view, and a hard-negative view. Three components are proposed to interplay and assist with each other.

    \item We experimentally and theoretically analyze the completeness of existing recommender models, the equivalence of BPR and \textit{entropy} on optimization, the mechanism of topological information, and the bounds of the hard negative sampling approach. 

    \item Extensive experiments demonstrate our diffusion framework's superiority, in terms of its performance, efficiency, and compatibility with diverse scenarios.
\end{itemize}

\section{Preliminaries}
\textit{Recommendation Task.} Given the data of user-item interaction $\mathcal{D}=\{u,i,r_{u,i}|u\in \mathcal{U},i\in \mathcal{I}\}$, $\mathcal{U}=\{u_1,\ldots,u_m\}$ denotes the user set ($|\mathcal{U}|=m$), $\mathcal{I}=\{i_1,\ldots,i_n\}$ denotes the item set ($|\mathcal{I}|=n$), and $r_{u,i}=\{0,1\}$ indicates whether user $u$ has interacted with item $i$ as implicit feedback. $N(u)$ is the interactive item set of user $u$. In matrix, the bipartite graph can be represented by $\textbf{\textit{R}}\in \{0,1\}^{m\times n}$. In our work, we focus on user-based Top-K recommendations, where K is the number of predicted items. We use \textbf{BOLD UPPERCASE} and \textbf{lowercase} letters to denote matrices and vectors, respectively.

\noindent\textit{Diffusion Model.} For a given data point $\textbf{\textit{x}}_0\in\mathbb{R}^n$ that represents a complete interaction record of a user (a.k.a. $\textbf{\textit{r}}_u$), the diffusion model~\cite{DiffRec} establishes a predefined variational distribution for the forward trajectory $q(\textbf{\textit{x}}_t|\textbf{\textit{x}}_{0})$ as:
\begin{align}
    % q(\textbf{\textit{x}}_t|\textbf{\textit{x}}_{t-1})&=\mathcal{N}(\textbf{\textit{x}}_t;\sqrt{1-\beta_t}\textbf{\textit{x}}_{t-1},\beta_t\textbf{\textit{I}}),\label{eq:q}\\
    q(\textbf{\textit{x}}_t|\textbf{\textit{x}}_{0})&=\mathcal{N}(\textbf{\textit{x}}_t;\prod_{t'=1}^t\sqrt{(1-\beta_{t'})}\textbf{\textit{x}}_0,(1-\prod_{t'=1}^t(1-\beta_{t'})\textbf{\textit{I}}), \label{eq:q0}
\end{align}
which is indexed by timesteps $t\in\{1,\ldots,T\}$. $\beta_t\in(0,1)$ denotes the magnitude of added noise on the timestep $t$. This trajectory progressively introduces incremental Gaussian noise $\mathcal{N}(\cdot)$, constructing a diffusion trajectory to corrupt $\textbf{\textit{x}}_{t-1}$ stepwisely. Consequently, the cumulative corruption gradually destroys original information, resulting in the final latent variable $\textbf{\textit{x}}_T$ that bears minimal resemblance to the original input $\textbf{\textit{x}}_0$.

The denoising part of the model counteracts the aforementioned process through learnable distributions $p_\theta(\textbf{\textit{x}}_{t-1}|\textbf{\textit{x}}_t)$ as:
\begin{equation}
    p_\theta(\textbf{\textit{x}}_{t-1}|\textbf{\textit{x}}_t)=\mathcal{N}(\textbf{\textit{x}}_{t-1};\boldsymbol{\mu}_\theta(\textbf{\textit{x}}_t,t),\boldsymbol{\Sigma}_\theta(\textbf{\textit{x}}_t,t)),
\end{equation}
where $\boldsymbol{\mu}_\theta(\cdot)$ and $\boldsymbol{\Sigma}_\theta(\cdot)$ denote mean and covariance of learnable distribution by parameter $\theta$, respectively. When the incremental noise added per diffusion step is sufficiently small, the denoising distributions can be sufficiently modeled as factorization across the data dimensions. 

For optimization, diffusion models leverage variational inference and KL divergences to calculate the tractable posterior on $\textbf{\textit{x}}_0$:
\begin{equation}
\begin{aligned}
    \log p(\textbf{\textit{x}}_0)\geq\mathbb{E}_q [\log p(\textbf{\textit{x}}_0|\textbf{\textit{x}}_1)-KL(q(\textbf{\textit{x}}_T|\textbf{\textit{x}}_0)\Vert p(\textbf{\textit{x}}_T)) \\[-1.3mm]
    -\sum^T_{t=2}KL(q(\textbf{\textit{x}}_{t-1}|\textbf{\textit{x}}_t,\textbf{\textit{x}}_0)\Vert p(\textbf{\textit{x}}_{t-1}|\textbf{\textit{x}}_t))],
\end{aligned}
\end{equation}
where $KL(q(\textbf{\textit{x}}_T|\textbf{\textit{x}}_0)\Vert p(\textbf{\textit{x}}_T))\approx0$ can be omitted \textit{iff.} the trajectory $q$ has been well-defined~\cite{thermoDiff}. For inference, diffusion models generate new data $\tilde{\textbf{\textit{x}}}_0$ from $\textbf{\textit{x}}_T$ iteratively by the trained distribution $p_\theta$.

\begin{figure*}[!t]
  \centering
  \includegraphics[width=1.\textwidth]{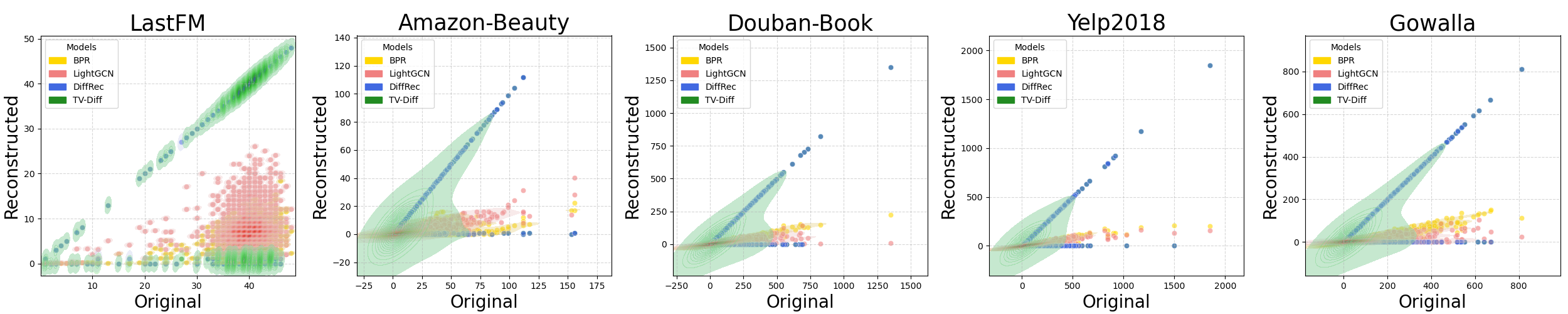}
  \includegraphics[width=1.\textwidth]{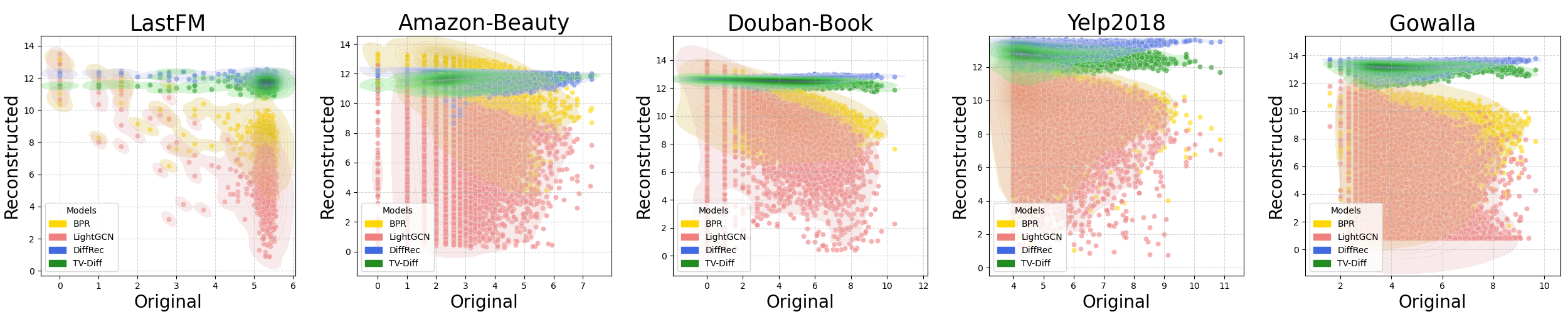}
  \vspace{-4mm}
  \caption{Visualization of \textit{Energy} (upper) and \textit{Entropy} (lower) on the user's original and reconstructed interaction probability.}
  \label{fig:visual}
  \vspace{-2mm}
\end{figure*}

\begin{figure}[!t]
  \centering
  \includegraphics[width=1.\columnwidth]{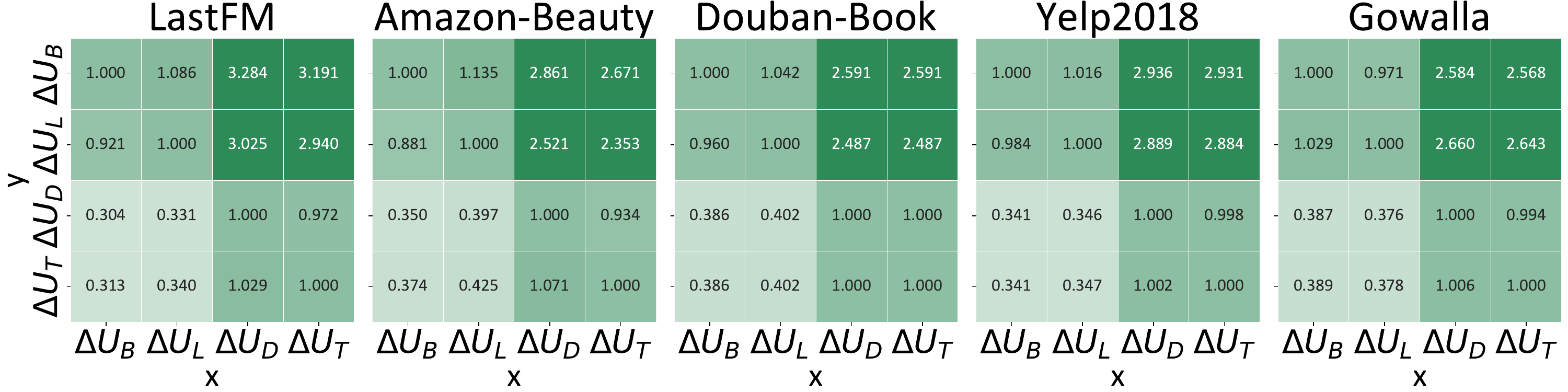}
  \includegraphics[width=1.\columnwidth]{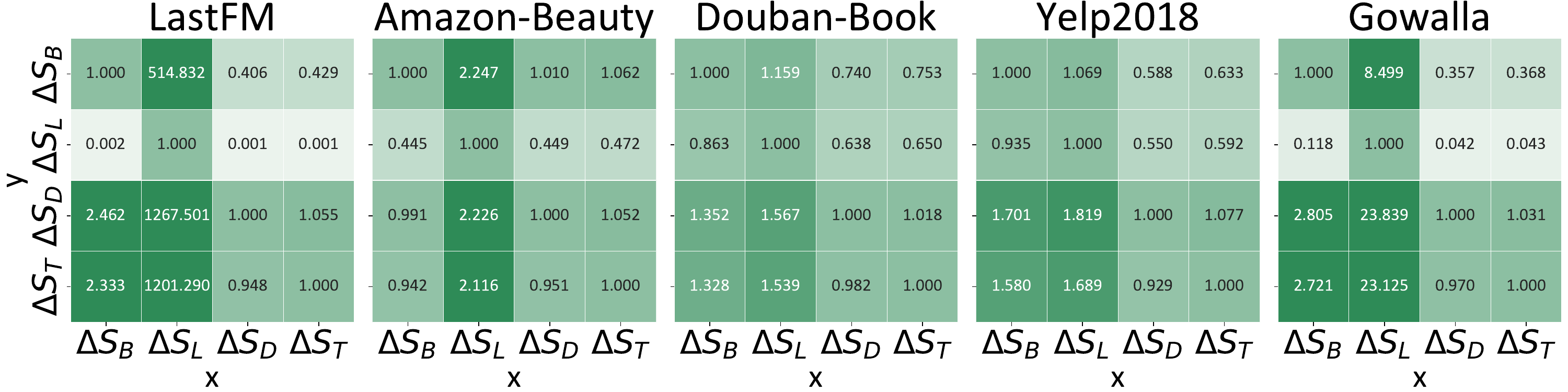}
  \vspace{-4mm}
  \caption{Detailed comparison among different models. $\Delta U_* \uparrow \in \mathbb{R}^{-}$ or $\Delta S_* \downarrow \in \mathbb{R}^{+}$ denotes the \textit{energy} or \textit{entropy} in difference on training, i.e., $U_*(\hat{\textbf{\textit{R}}'})-U_*(\hat{\textbf{\textit{R}}})$ or $S_*(\hat{\textbf{\textit{R}}'})-S_*(\hat{\textbf{\textit{R}}})$. The element in blocks denotes the $\frac{y}{x}$.}
  \label{fig:delta}
  \vspace{-2mm}
\end{figure}

\section{Tri-View Diffusion Recommendation Framework}
\subsection{Thermodynamic View - Helmholtz Free Energy Maximization}
To investigate the completeness of DM-based recommender models and classic recommender models from a thermodynamic view, we choose three representative baselines, i.e., BPR\cite{BPR}, LightGCN~\cite{LightGCN}, and DiffRec\cite{DiffRec}, to visualize the original users' interaction probability $\hat{\textbf{\textit{R}}}$ and the reconstructed ones $\hat{\textbf{\textit{R}}'}$ after training :
\begin{gather}
    \hat{\textbf{\textit{R}}}=\textbf{\textit{D}}_{\mathcal{U}}^{-1}\textbf{\textit{R}},\\
    \hat{\textbf{\textit{R}}'} = \begin{cases} \tilde{\textbf{\textit{D}}}_{\mathcal{U}}^{-1}\tilde{\textbf{\textit{R}}}, &\mbox{DiffRec}\\
    Softmax(\textbf{\textit{E}}_{\mathcal{U}}\cdot \textbf{\textit{E}}_{\mathcal{I}}^{\top}), &\mbox{BPR, LightGCN}
    \end{cases},
\end{gather}
where $\tilde{\textbf{\textit{R}}}\in\mathbb{R}^{m\times n}$ denotes the raw reconstruction (a.k.a. \textit{scores}), $\textbf{\textit{D}}_{\mathcal{U}}\in \mathbb{R}^{m\times m}$ and $\tilde{\textbf{\textit{D}}}_{\mathcal{U}}=diag(\frac{1}{\Vert\tilde{\textbf{\textit{r}}}_1\Vert_1},\ldots,\frac{1}{\Vert\tilde{\textbf{\textit{r}}}_m\Vert_1})$ denotes the original / reconstructed user degree matrix, $\Vert\cdot\Vert_1$ denotes the L1-norm, $\textbf{\textit{E}}_{\mathcal{U}}\in \mathbb{R}^{m\times d}$ and $\textbf{\textit{E}}_{\mathcal{I}}\in \mathbb{R}^{n\times d}$ are user / item representations, $Softmax(\cdot)$ denotes the row-wise softmax function. For clarity, we only take the acronym of model names hereafter. Note that this preprocessing step is implemented following their original prediction procedures, aiming to normalize users' interaction probabilities as a prerequisite for subsequent measurement. In terms of \textit{energy} in thermodynamics, one of the most commonly used metrics is to measure the proportional number of interactions~\cite{energy}, so-called \textit{explicit energy}:
\vspace{-0.6\baselineskip}
\begin{equation}
    % =tr(\hat{\textbf{\textit{A}}'}\hat{\textbf{\textit{A}}'}^{\top})^{\frac{1}{2}} 
    % \propto tr(\textbf{\textit{D}}_{\mathcal{U}})
    U(\hat{\textbf{\textit{R}}'})=tr(|\hat{\textbf{\textit{R}}'}|) \propto tr(\textbf{\textit{D}}_{\mathcal{U}}), \quad \hat{r}'_{u,i} :=\begin{cases}
    1, \quad \hat{r}'_{u,i} \geq \hat{r}_{u,i}\\[0.1mm]
    0, \quad otherwise %\hat{a}'_{u,i} < \hat{a}_{u,i}
    \end{cases},
\end{equation}
where $tr(\cdot)$ denotes the trace of a matrix. $| \hat{\textbf{\textit{R}}'}| :=(\hat{\textbf{\textit{R}}'}\hat{\textbf{\textit{R}}'}^{*})^{\frac{1}{2}}$ where $\hat{\textbf{\textit{R}}'}^{*}$ is its conjugate transpose. However, such a strict classification would inevitably overlook those reconstructions that marginally fail the threshold, i.e., \textit{implicit energy}. Hence, we fix the definition to involve both of them comprehensively:
\begin{equation}
    U(\hat{\textbf{\textit{R}}'}) = \sum_u\sum_i\mathbb{I}(\hat{r}'_{u,i} \geq r_{u,i}) + \mathbb{I}(\hat{r}'_{u,i} < r_{u,i})\frac{\hat{r}'_{u,i}}{\hat{r}_{u,i}},\label{eq:7}
\end{equation}
where $\mathbb{I}(\cdot)=\{0,1\}$ denotes the indication function. Specifically, the \textit{explicit energy} refers to the potential links that are higher than the original probability, and the \textit{implicit energy} means the weighted distance-based affinity. On the other hand, the equation of \textit{entropy} is defined by \textit{Shannon Entropy} for simplicity~\cite{entropy}:
\begin{equation}
    S(\hat{\textbf{\textit{R}}'}) = \sum_u -\hat{\textbf{\textit{r}}}'_u \log \hat{\textbf{\textit{r}}}'^{\top}_u.
\end{equation}

According to results on five datasets (cf. Figure~\ref{fig:visual} \& Figure~\ref{fig:delta}), we draw the following obvervations:

\noindent\textsc{Observation 1.} \label{observation:1} \textit{DM-based recommender models always preserves \ul{higher} \textit{energy} than the classic ones, i.e., $\frac{\Delta U_D}{\Delta U_{\{B,L\}}}\in(0,1)$.}

It shows the connection between \textit{energy} and the loss of diffusion models, which is commonly defined by the Mean Squared Error (MSE). By optimization, the reconstructed interaction matrix resembles the original one, with the increment of \textit{energy}.

\textsc{Observation 2.} \label{observation:2} \textit{DM-based recommender models always fail to achieve \ul{lower} \textit{entropy} than the classic ones, i.e., $\frac{\Delta S_D}{\Delta S_{\{B,L\}}} > 1$.Specifically, DM-based recommender models are almost \textit{isentropic}.}

It shows the equivalence between \textit{entropy} and the loss of classic recommender models, where BPR is employed in general. See the theoretical proof in Appendix~\ref{proof:BPR} for details. However, none of the objectives in DM-based recommender models reflect the \textit{entropy}.

\textsc{Observation 3.} \label{observation:3} \textit{Graph-based recommender model (i.e., LightGCN) exhibits \ul{stronger} capability on \textit{entropy}-reduction than its backbone (i.e., BPR), i.e., $\frac{\Delta S_B}{\Delta S_L} > 1$.}

It attributes the significant decrease of \textit{entropy} to the neighbors' information during message-passing. This effect is relevant to topological and anisotropic signals on the user-item bipartite graph, which will be discussed in detail in Section~\ref{subsection:topology}.

To sum up, DM-based recommender models and classic recommender models follow two independent optimization schemes, which reveal potential for a complete enhancement. To this end, we propose the \textit{Helmholtz Free Energy} Maximization:
\begin{align}
    \label{eq:helmholtz}
    \mathcal{L}_H(\hat{\textbf{\textit{R}}'}) \equiv &\ \mathcal{L}_U(\hat{\textbf{\textit{R}}'}) - t \cdot \mathcal{L}_S(\hat{\textbf{\textit{R}}'})  = U(\hat{\textbf{\textit{R}}'}) - t\cdot S(\hat{\textbf{\textit{R}}'}) \nonumber\\
    =&\ -\sum_{u\in\mathcal{B}}\sum_i(\hat{r}_{u,i}-\hat{r}'_{u,i})^2 \\ 
    &\ - t\cdot 
    (\hat{r}_{u,i}\log \sigma(\hat{r}'_{u,i})-c_{u,i}\cdot(1-\hat{r}_{u,i})\log(1-\sigma(\hat{r}'_{u,i}))), \nonumber
\end{align}
where $\mathcal{B}$ denotes the user batch for training, $t$ denotes the temperature, a coefficient to balance the priority of the two metrics. $\sigma(\cdot)$ denotes the sigmoid function. $c_{u,i}\in\{0,1\}$ denotes the confidence to sample the non-interactive items (See Section~\ref{subsection:NS}). The \textit{Helmholtz free energy} expresses the maximum useful work extractable from a thermodynamic system given constant temperature. It reaches towards an equilibrium by maximizing \textit{energy} and minimizing \textit{entropy} in the meantime. Through the maximization of \textit{free energy}, DM-based recommender models avert the collapse to suboptimal outcomes where \textit{entropy} remains constant, leading to more generalized and robust results for personalization recommendations. It is worth noting that, Binary Cross Entropy (BCE) is intuitively adopted as the \textit{entropy-based} loss due to its consistency with the user-wise computation on the \textit{energy-based} loss. Nonetheless, alternative choices are also available, e.g., BPR, Negative Log Likelihood (NLL) that neglects the non-interactive class (i.e., $\hat{r}_{u,i}=0$). We will compare their performance in Section~\ref{subsection:ablation}. 

\subsection{Topological View - Anisotropic Denoiser}
\label{subsection:topology}
Based on Observation 3 in Section~\ref{observation:3}, we acknowledge that topological information, or graph structure, expresses rich relationships between users and items. The local and global neighborhoods are leveraged to exploit the connectivity patterns, e.g., anisotropy, centrality, and communities, for personalized recommendations. Specifically, the user-item bipartite graph is represented as $\mathcal{G}=(\mathcal{V},\mathcal{E})$, where $\mathcal{V}=\mathcal{U}\cup \mathcal{I}$ denotes all user and item nodes and $\mathcal{E}$ denotes all the observed interactions between users and items. The symmetric normalized bipartite matrix without self-connection is defined as $\bar{\textbf{\textit{R}}} = \textbf{\textit{D}}_{\mathcal{U}}^{-\frac{1}{2}}\textbf{\textit{R}}\textbf{\textit{D}}_{\mathcal{I}}^{-\frac{1}{2}}$. Although topological information has been roughly adopted to DM-based recommender models, how the underlying anisotropy within the bipartite topology benefit to the optimization process of DM-based recommender models remain unclear. In the context of \textit{entropy}, we first arrive at a theorem that articulates its connection with topological information.

\begin{lemma}
    \vspace{-0.3\baselineskip}
    \label{lemma:BPR}
    Given a model trained on \textit{entropy} (i.e., BPR), the optimal results approximate the bipartite matrix \textbf{\textit{R}}.
\end{lemma}
\vspace{-0.3\baselineskip}

\begin{proof}
    \vspace{-0.5\baselineskip}
    The goal for BPR is to maximize the score estimator $s_{u,i+,i-}:=s_{u,i+}-s_{u,i-}$ for all user-positive-negative triplets ($u,i+,i-$). When $s_{u,i+,i-}\to +\infty$, $s_{u,i+}\to +\infty$ and $s_{u,i-}\to -\infty$. After the activation function $\tilde{r}_{u,*}=\sigma(s_{u,*})=1/(1+\exp^{-s_{u,*}})$, the optimal results are $\tilde{r}_{u,i+}=1$ and $\tilde{r}_{u,i-}=0$, which is the original binary interaction graph matrix $\textbf{\textit{R}}$.
    \pushQED{}
\end{proof}
\vspace{-0.3\baselineskip}

\begin{theorem}
    \vspace{-0.5\baselineskip}
    \label{theorem:graph}
    Given a model trained on \textit{entropy} (i.e., BPR), graph-based recommender models (i.e., LightGCN) with graph filter $\textbf{\textit{D}}_{\mathcal{U}}^{-\frac{1}{2}}\textbf{\textit{R}}\textbf{\textit{D}}_{\mathcal{I}}^{-\frac{1}{2}}$ decreases more entropy during optimization, i.e., $\Delta S(\hat{\textbf{\textit{R}}}_L|\hat{\textbf{\textit{R}}}_B)\leq0$. 
\end{theorem}
\vspace{-0.3\baselineskip}

\begin{proof}
    \vspace{-0.5\baselineskip}
    According to Lemma~\ref{lemma:BPR}, the probabilistic matrix of BPR is $\hat{\textbf{\textit{R}}}_B=\textbf{\textit{D}}_\mathcal{U}^{-1}\textbf{\textit{R}}=[r_{u,i}/d_u]_{m\times n}=[\hat{r}_{u,i}]_{m\times n}$, where $d_u$ is the degree of user $u$. Similarly, LightGCN learns the graph structures by training, such that the prediction results approach to  $\tilde{\textbf{\textit{R}}}_{L} = \textbf{\textit{D}}_{\mathcal{U}}^{-\frac{1}{2}}\textbf{\textit{R}}\textbf{\textit{D}}_{\mathcal{I}}^{-\frac{1}{2}}$. After normalization for \textit{entropy} calculation, the probabilistic matrix of LightGCN is $\hat{\textbf{\textit{R}}}_L=\tilde{\textbf{\textit{D}}}_{\mathcal{U}}^{-1}\tilde{\textbf{\textit{R}}}_L=[r_{u,i}/(\sqrt{d_u}\sqrt{d_i}\sum_j(r_{u,i}/(\sqrt{d_u}\sqrt{d_j})))]_{m\times n}$ $=[\hat{r}'_{u,i}]_{m\times n}$, where $d_i$ and $d_j$ denotes the degree of item $i$ and $j$, respectively. Hence, the pivotal point of proof lies in the calculation of the difference between $S(\hat{\textbf{\textit{R}}}_B)$ and $S(\hat{\textbf{\textit{R}}}_L)$:
    \begin{equation}
    \begin{split}
        \Delta S(\hat{\textbf{\textit{R}}}_L|\hat{\textbf{\textit{R}}}_B) &= S(\hat{\textbf{\textit{R}}}_L)-S(\hat{\textbf{\textit{R}}}_B)\\
        &=\sum_u\sum_i- \hat{r}'_{u,i}\log\hat{r}'_{u,i} + \hat{r}_{u,i}\log\hat{r}_{u,i}.
    \end{split}
    \end{equation}
    After simplifying the equation, we have:
    \begin{align}
        \Delta S(\hat{\textbf{\textit{R}}}_L|\hat{\textbf{\textit{R}}}_B) =\sum_u \sum_i -\overbrace{\frac{r_{u,i}}{\sqrt{d_i}\sum_j(r_{u,j}/\sqrt{d_j})}\log r_{u,i}}^{\mbox{\color{orange}=0}} \label{eq:11}\\
        + \frac{r_{u,i}}{\sqrt{d_i}\sum_j(r_{u,j}/\sqrt{d_j})}\log ({\sqrt{d_i}\sum_j(r_{u,j}/\sqrt{d_j})}) &- \frac{r_{u,i}}{d_u}\log d_u \nonumber
    \end{align}
    \vspace{-2mm}
    \begin{align}
        =\sum_u &\sum_i r_{u,i}(\frac{\log(\sqrt{d_i}\sum_j (r_{u,j}/\sqrt{d_j})}{\sqrt{d_i}\sum_j (r_{u,j}/\sqrt{d_j})}-\frac{\log d_u}{d_u})\nonumber\\
        =\sum_u &(\Vert [-\frac{1}{\sqrt{d_i}\sum_j(r_{u,j}/\sqrt{d_j})}\log \frac{1}{\sqrt{d_i}\sum_j(r_{u,j}/\sqrt{d_j})}]_{d_u} \Vert_1\nonumber\\
        -& \underbrace{\Vert [-\frac{1}{d_u}\log \frac{1}{d_u}]_{d_u}\Vert_1}_{\color{orange}\mbox{Max.}}) \leq0\qquad \text{subject to}\ \forall i\in N(u) \label{eq:12},
    \end{align}
    where $\Vert\cdot\Vert_1$ denotes L1-norm. The first term of the primary equation (cf. Eq. (\ref{eq:11})) becomes 0 for $r_{u,*}\in {\{0,1\}}$. Since the second term of the final equation (cf. Eq. (\ref{eq:12})) achieves the maximum \textit{entropy} for interactive-item vectors $[\cdot]_{d_u}$, the \textit{entropy} difference between the two models is always non-positive. Hence, though as a low-pass filter, $\bar{\textbf{\textit{R}}}$ attains \ul{\textit{NO} smoother} item probability counterintuitively.
    % the final result $\Delta S(\hat{\textbf{\textit{A}}}_L|\hat{\textbf{\textit{A}}}_B)\leq0$ is well proved.
    \label{proof:entropy}
    \pushQED{}
\end{proof}
\vspace{-0.5\baselineskip}

\begin{remark}
    \vspace{-0.4\baselineskip}
    In general, \textit{energy} and \textit{entropy} act as countervailing forces, akin to DM-based versus classic recommender models. However, though Theorem~\ref{theorem:graph} posits lower \textit{entropy} for LightGCN, its implementation of multi-layer message-passing potentially explores non-interactive items. Thus, LightGCN might paradoxically result in higher overall \textit{energy} (cf. Figure~\ref{fig:delta}). Noting that, message-passing with more layers inevitably increases entropy (See Appendix~\ref{proof:MP}), necessitating a trade-off consideration on the total number of layers.
    % Theorem~\ref{theorem:graph} also provides proof that LightGCN can theoretically achieve more \textit{full-implicit energy}, i.e., $\textit{U}(\hat{\textbf{\textit{A}}'})=\sum_u\sum_i\frac{\hat{a}'_{u,i}}{\hat{a}_{u,i}}$ and $\Delta U(\hat{\textbf{\textit{A}}}_L|$ $\hat{\textbf{\textit{A}}}_B)\geq0$. Nevertheless, simply measuring implicit energy will incur energy overload, where the reconstructed energy is higher than the original, leading to disorderly results. Thus, by considering both explicit and implicit energy via the non-continuous indication function $\mathbb{I}(\cdot)$ (cf. Eq.~\ref{eq:7}), the energy of LighGCN is usually higher than BPR, yet exceptions may exist (e.g., Gowalla in Figure~\ref{fig:delta}).
\end{remark}
\vspace{-0.3\baselineskip}

To sum up, incorporating item degree information into the topology facilitates \textit{entropy}-based training objective to achieve higher \textit{Helmholtz free energy} via \textit{entropy} reduction, eventually improving model performance. Furthermore, the inconsistency between user and item degree distribution, which is the exclusive anisotropy inherent to user-item bipartite graphs, necessitates an asynchronous modeling. To this end, we propose the anisotropic denoiser: 
\begin{align}
    \begin{split}
        \tilde{\textbf{\textit{x}}}_\theta(\textbf{\textit{x}}_t, t) &= {\textbf{\textit{h}}}_{u|\theta}(\textbf{\textit{x}}_t, t) \cdot {\textbf{\textit{H}}}_{\mathcal{I}|\theta}(\textbf{\textit{x}}_t,t)^{\top}\\
        &= \tanh(Agg(\textbf{\textit{x}}_t \cdot \textbf{\textit{W}}_{\mathcal{I}}, \textbf{\textit{e}}_t)) \cdot \tanh(\bar{\textbf{\textit{R}}}^{\top}\cdot\textbf{\textit{W}}_{\mathcal{U}})^{\top},
        \label{eq:denoise1}
    \end{split}\\
    \hat{\textbf{\textit{r}}}'_{u} &= \tilde{\textbf{\textit{x}}}_\theta(\textbf{\textit{x}}_t, t)/\Vert \tilde{\textbf{\textit{x}}}_\theta(\textbf{\textit{x}}_t, t)\Vert_1,
    \label{eq:denoise2}
\end{align}
where $\textbf{\textit{W}}_{\mathcal{I}}\in\mathbb{R}^{n\times d}$, $\textbf{\textit{W}}_{\mathcal{U}}\in\mathbb{R}^{m\times d}$ denotes the latent transformation matrices, $\textbf{\textit{e}}_{t}\in \mathbb{R}^{d}$ denotes the learnable time embedding. $\bar{\textbf{\textit{R}}}\in \mathbb{R}^{m\times n}$ denotes the symmetric normalized bipartite matrix. $Agg(\cdot)$ denotes the aggregation function, e.g., element-wise addition. $\tanh(\cdot)$ is the activation functions as~\cite{DiffRec}. In conclusion, the user-wise encoder-decoder paradigm in existing DM-based approaches solely models the user degree distribution and presumes the item degree distribution in line with it. In contrast, the anisotropic denoiser explicitly accounts for the item degree on explicit topological information and drives predictions by calculating the user-item cross-correlation. According to Theorem~\ref{proof:entropy}, we only involve single-layer message-passing to preserve anisotropic signals and avoid loss of \textit{entropy} from multi-layer message-passing. It also saves computational resources, particularly on pre-training or multi-layer message-passing on large graphs constructed from large datasets. Consequently, TV-Diff can obviate complex and sometimes redundant efforts in capturing anisotropic signals. 

\subsection{Hard-Negative View - Acceptance-Rejection Gumbel Sampling Process}
\label{subsection:NS}
% The long-tail distribution~\cite{LongTail} of items and inherent data sparsity in recommendation tasks have motivated increasing attention to negative sampling, a strategy that utilizes patterns within non-interactive items. 
The maximization of \textit{Helmholtz free energy} integrates both genres of objectives for optimization. It necessitates an efficient yet effective negative sampling method to ensure the consistent robustness of their mutual joint-training. However,
the abundance of non-interactive items overwhelms the distribution of positive items when the model is trained on the entire itemset, deteriorating the model's efficacy and efficiency. The confidence $c_{u,i}\sim Bernoulli(\frac{N(u)}{n})$ is thus introduced to balance the far-outnumbered non-interactive items, through which the model randomly samples negative items at a rate proportional to the number of interactions~\cite{BPR}. Although hard negative sampling (Hard NS) methods~\cite{MCNS,DMNS} retrieve high-quality informative negative items, capture informative semantics, and accelerate convergence during training, they impose the negative distribution $p_n(j|u)$ sub-linearly correlated to the positive $p_d(i|u)$, are not judicious for the recommendation tasks on the bipartite interaction graph. By corroborating this understanding, we delve into the negative distribution on one of the simplest recommender models (i.e., BPR) with its score estimator $s_{u,*}$:

\begin{theorem}
    \vspace{-0.3\baselineskip}
    The expected loss of BPR reaches the upper bound if and only if $p_n(j|u)=\mathcal{C}\cdot p_d(i|u)$, and the lower bound if $p_n(j|u)\cdot p_d(i|u)=0$ for all positive-negative item pair $(i,j)$.
\end{theorem}
\vspace{-0.5\baselineskip}

\begin{proof}
    \vspace{-0.5\baselineskip}
    The objective function for each user $u$ is
    \begin{equation}
        \begin{split}
        \mathcal{L}_{BPR}{(u)}&=\mathbb{E}_{i_{\textasciitilde} p_d(u)}\mathbb{E}_{j_{\textasciitilde} p_n(u)}[-\log \sigma(\textbf{\textit{e}}_{u}\textbf{\textit{e}}_{i}^{\top}-\textbf{\textit{e}}_{u}\textbf{\textit{e}}_{j}^{\top})]\\
        &=\sum_i p_d(i|u) \sum_jp_n(j|u)[-\log\sigma({s}_{u,i}-{s}_{u,j})]\\
        &\leq \sum_i\sum_j [-\log\sigma({s}_{u,i}-{s}_{u,j})],
        \label{eq:NS}
        \end{split}
    \end{equation}
    where $\sum_i p_d(i|u)=\sum_j p_n(j|u)=1$, and the upper bound (cf. Eq. (\ref{eq:NS})) holds according to \textit{Cauchy–Schwarz Inequality}. The maximizer satisfies $p_n(j|u)=\mathcal{C}\cdot\mathbb{E}[-\log\sigma({s}_{u,i}-{s}_{u,j})]$, where $\mathcal{C}$ denotes the constant. Conversely, the minimizer of $\mathcal{L}_{BPR}{(u)}$ arrives at 0 if it adheres to the condition that they are orthogonal. The same applies to $p_d(i|u)$. Therefore, when the positive and negative distributions are correlated in a sub-linear manner, BPR loss will compromise amid optimization (orthogonality) - generalization (linearity). The same issue arises in BCE due to their equivalence to \textit{entropy}-based losses. In contrast to previous work that imposes the negative and positive distribution sub-linearly correlated, enforcing orthogonality between them preserves informative semantics (e.g., anisotropy) for the training of the denoiser.
    \label{proof:NS}
    \pushQED{}
\end{proof}
\vspace{-0.5\baselineskip}
\vspace{-0.5\baselineskip}
To properly sample hard negatives deviated from positives, we adopt the acceptance-rejection sampling~\cite{RejectionSampling} to truncate the potentially non-informative items:
\begin{align}
    c_{u,j}\sim p_n(j|u)
    \approx \begin{cases}
        \frac{1}{\gamma \cdot n}, &\mbox{Rank}(\tilde{r}_{u,j})\leq\gamma\cdot n\\[3pt]
        \epsilon, &\mbox{Rank}(\tilde{r}_{u,j})>\gamma \cdot n
    \end{cases},
    \label{eq:AR}
\end{align}
where $\gamma\in(0, 1]$ denotes the negative factor controlling the threshold of informative negatives, $\epsilon$ denotes the infinitesimal, and $\mbox{Rank}(\cdot)$ denotes the function that returns the position of item $j$ in the monotonically decreasing-order list of all items scores $\tilde{r}_{u,*}\propto s_{u,*}$. 
However, in DM-based recommender models, the scores can be unreliable: When large diffusion timesteps are sampled, the forward process will overcorrupt the input. It severely challenges the denoiser's capability in the reverse process, barely reconstructing the scores accurately. As a result, the Hard-NS method struggles in distinguishing negatives and degrades model performance due to false negatives. To address this issue, we apply a timestep-dependent Gumbel Softmax on top of acceptance-rejection sampling:
\begin{equation}
\begin{split}
    \hat{p}_n(j|u)&=Softmax(\frac{\log p_n(j|u)+g_j}{\tau(\bar{t})})
    =\frac{\exp(\frac{\log p_n(j|u)+g_j}{\exp(-\lambda \bar{t})})}{\sum_{j'}\exp(\frac{\log p_n(j'|u)+g_{j'}}{\exp(-\lambda \bar{t})})},
    \label{eq:GSP}
\end{split}
\end{equation}
where $g\sim Gumbel(0,1)$ is sampled by the reparameterization trick, i.e., $g=-\log(\log(\mu)), \mu\sim Uniform(0,1)$. $\lambda$ denotes the relaxation rate of Hard NS. $\bar{t}=1-\frac{t}{T}$ ensures that when $t\to T$, $\tau(\bar{t})\to 1$ reaches the maximum, and the negative distribution towards a uniform distribution among all items to avoid mistakes of negatives.
% On the other hand, when $t\to0$, the distribution becomes more concentrated at the previous. 
Altogether, the Acceptance-Rejection Gumbel Sampling Process (AR-GSP) establishes a dual-aspect Hard-NS strategy: On the general recommendation aspect, acceptance-rejection sampling manageably dissects the item pool into positives and long-tailed true negatives~\cite{LongTail}. Then, true positives and false positives (i.e., hard negatives) contribute evenly to the model's generalization. On the diffusion aspect, the time-dependent Gumbel-Softmax is a DM-specific approach that assesses the reliability of the preliminary negative distribution and adaptively provides a viable one.

\begin{table*}[t]\scriptsize
    \tabcolsep=0.05cm
    \caption{Overall performance comparison with representative models on five datasets. For acronyms used in Type, "B" represents base recommenders; "A" refers to autoencoder-based recommenders; "G" means the graph-based recommenders; "N" means the negative sampling recommenders; "D" represents diffusion-based recommenders.}
    \vspace{-2mm}
    \begin{tabular*}{\textwidth}{@{\extracolsep{\fill}}c|c|cccc|cccc|cccc|cccc|cccc}
        \cmidrule{1-22}
        \multicolumn{2}{c}{\textbf{Dataset}} & \multicolumn{4}{c}{\textbf{LastFM}} & \multicolumn{4}{c}{\textbf{Amazon-Beauty}} & \multicolumn{4}{c}{\textbf{Douban-Book}} & \multicolumn{4}{c}{\textbf{Yelp2018}} & \multicolumn{4}{c}{\textbf{Gowalla}}\\ 
        \cmidrule(lr){1-2} \cmidrule(lr){3-6} \cmidrule(lr){7-10} \cmidrule(lr){11-14} \cmidrule(lr){15-18} \cmidrule(lr){19-22}
        
        \textbf{Type} & \textbf{Method} & \textbf{R@10} & \textbf{N@10} & \textbf{R@20} & \textbf{N@20} & \textbf{R@10} & \textbf{N@10} & \textbf{R@20} & \textbf{N@20} & \textbf{R@10} & \textbf{N@10} & \textbf{R@20} & \textbf{N@20} & \textbf{R@10} & \textbf{N@10} & \textbf{R@20} & \textbf{N@20} & \textbf{R@10} & \textbf{N@10} & \textbf{R@20} & \textbf{N@20}\\ \cmidrule{1-22} 
            
        \multirow{2}{*}{\textbf{B}} 
        & BPR-MF & 0.1742 & 0.2070 & 0.2602 & 0.2435 & 0.0711 & 0.0470 & 0.1015 & 0.0572 & 0.0776 & 0.0927 & 0.1268 & 0.1038 & 0.0279 & 0.0316 & 0.0491 & 0.0397 & 0.1045 & 0.1052 & 0.0456 & 0.1213 \\[0.1em]
        & NeuMF & 0.1454 & 0.1669 & 0.2257 & 0.2056 & 0.0622 & 0.0397 & 0.0937 & 0.0496 & 0.0572 & 0.0673 & 0.0895 & 0.0735 & 0.0224 & 0.0250 & 0.0396 & 0.0314 & 0.0932 & 0.0904 & 0.1398 & 0.1051 \\
        \cmidrule{1-22}

        \multirow{2}{*}{\textbf{A}} 
        & CDAE & 0.0752 & 0.0877 & 0.1082 & 0.0906 & 0.0208 & 0.0119 & 0.0341 & 0.0157 & 0.0463 & 0.0508 & 0.0716 & 0.0579 & 0.0073 & 0.0081 & 0.0127 & 0.0102 & 0.0209 & 0.0190 & 0.0305 & 0.0210\\[0.1em]
        & MultiVAE & 0.1040 & 0.1220 & 0.2713 & 0.2488 & 0.0576 & 0.0391 & 0.1018 & 0.0574 & 0.0459 & 0.0475 & 0.1207 & 0.1008 & 0.0245 & 0.0271 & 0.0542 & 0.0434 & 0.0763 & 0.0711 & 0.1480 & 0.1121\\
        \cmidrule{1-22}

        \multirow{4.5}{*}{\textbf{G}} 
        & LightGCN & 0.1980 & 0.2347 & 0.2999 & 0.2836 & 0.0921 & 0.0596 & 0.1308 & 0.0735 & 0.1000 & 0.1161 & 0.1504 & 0.1264 & 0.0342 & 0.0390 & 0.0587 & 0.0482 & 0.1278 & 0.1297 & 0.1857 & 0.1474\\[0.1em]
        & ChebyCF & 0.1868 & 0.2261 & 0.2981 & 0.2853 & 0.0839 & 0.0595 & 0.1358 & 0.0795 & 0.1107 & 0.1245 & 0.1762 & 0.1644 & 0.0348 & 0.0398 & 0.0598 & 0.0495 & 0.1318 & 0.1323 & 0.2002 & 0.1577\\[0.1em]
        & LinkProp & 0.1182 & 0.1446 & 0.2763 & 0.2647 & 0.0884 & 0.0648 & 0.1202 & 0.0747 & 0.0929 & 0.1088 & 0.1727 & 0.1513  & 0.0352 & 0.0402 & 0.0635 & 0.0517 & 0.1263 & 0.1261 & 0.2084 & 0.1642\\[0.1em]
        & SGCL & 0.1899 & 0.2305 & \ul{0.3093} & \ul{0.2946} & \ul{0.1004} & \ul{0.0678} & \ul{0.1433} & \ul{0.0810} & \ul{0.1405} & \ul{0.1704} & \ul{0.1939} & \ul{0.1771} & \textbf{0.0416} & \ul{0.0472} & \ul{0.0679} & \ul{0.0555} & \ul{0.1498} & \ul{0.1516} & \ul{0.2172} & \ul{0.1705}\\
        \cmidrule{1-22}

        \multirow{2}{*}{\textbf{N}} 
        & MixGCF & 0.1935 & 0.2306 & 0.3009 & 0.2838 & 0.0933 & 0.0607 & 0.1356 & 0.0754 & 0.0940 & 0.1104 & 0.1502 & 0.1242 & 0.0334 & 0.0378 & 0.0621 & 0.0509 & 0.1248 & 0.1254 & 0.1959 & 0.1542\\[0.1em]
        & AHNS & 0.1921 & 0.2285 & 0.2744 & 0.2611 & 0.0905 & 0.0583 & 0.1427 & 0.0802 & 0.1285 & 0.1633 & 0.1839 & 0.1690 & 0.0366 & 0.0422 & 0.0631 & 0.0519 & 0.1160 & 0.1183 & 0.1724 & 0.1346\\
        \cmidrule{1-22}

        \multirow{7}{*}{\textbf{D}} 
        & CODIGEM & 0.2122 & 0.2530 & 0.2943 & 0.2831 & 0.0666 & 0.0478 & 0.0916 & 0.0556 & 0.1115 & 0.1454 & 0.0716 & 0.0563 & 0.0349 & 0.0407 & 0.0590 & 0.0491 & 0.0930 & 0.0997 & 0.1347 & 0.1099\\[0.1em]
        & DiffRec & \ul{0.2127} & \ul{0.2538} & 0.2985 & 0.2867 & 0.0850 & 0.0586 & 0.1206 & 0.0713 & 0.1356 & 0.1682 & 0.1847 & 0.1695 & 0.0380 & 0.0443 & 0.0603 & 0.0506 & 0.1383 & 0.1439 & 0.1981 & 0.1593\\[0.1em]
        & BSPM & 0.1962 & 0.2349 & 0.2852 & 0.2708 & 0.0814 & 0.0602 & 0.1405 & 0.0797 & 0.1296 & 0.1594 & 0.1843 & 0.1740 & \ul{0.0401} & 0.0455 & 0.0670 & 0.0553 & 0.1484 & 0.1498 & 0.2023 & 0.1577\\[0.1em]
        & GiffCF & 0.1735 & 0.2112 & 0.2582 & 0.2501 & 0.0673 & 0.0488 & 0.1019 & 0.0620 & 0.0502 & 0.0578 & 0.1586 & 0.1344 & 0.0327 & 0.0394 & 0.0642 & 0.0529 & 0.1371 & 0.1397 & 0.1885 & 0.1470\\[0.1em]
        & DDRM & 0.1966 & 0.2337 & 0.2815 & 0.2675 & 0.0949 & 0.0627 & 0.1365 & 0.0760 & 0.0830 & 0.0957 & 0.1272 & 0.1065 & 0.0328 & 0.0377 & 0.0556 & 0.0457 & 0.1147 & 0.1163 & 0.1674 & 0.1317\\[0.1em]
        & HDRM & 0.1807 & 0.2139 & 0.2951 & 0.2782 & 0.0667 & 0.0438 & 0.1300 & 0.0695 & 0.0769 & 0.0914 & 0.1281 & 0.1080 & 0.0217 & 0.0257 & 0.0404 & 0.0332 & 0.0800 & 0.0773 & 0.1293 & 0.0959\\
        \cmidrule{1-22}
        \rowcolor{gray!20}
        \multirow{3}{*}{\textbf{-}} 
        & \textbf{TV-Diff} & \textbf{0.2220} & \textbf{0.2605} & \textbf{0.3133} & \textbf{0.2953} & \textbf{0.1046} & \textbf{0.0719} & \textbf{0.1475} & \textbf{0.0852} & \textbf{0.1474} & \textbf{0.1863} & \textbf{0.2037} & \textbf{0.1908} & \textbf{0.0416} & \textbf{0.0481} & \textbf{0.0700} & \textbf{0.0582} & \textbf{0.1517} & \textbf{0.1578} & \textbf{0.2195} & \textbf{0.1762}\\[0.1em]
        & \textbf{Imp. (Follow-up)} & \textbf{4.37\%} & \textbf{2.64\%} & \textbf{1.29\%} & \textbf{0.24\%} & \textbf{4.18\%} & \textbf{6.05\%} & \textbf{2.93\%} & \textbf{5.19\%} & \textbf{4.91\%} & \textbf{9.33\%} & \textbf{5.05\%} & \textbf{7.74\%} & {0.00\%} & \textbf{1.91\%} & \textbf{3.09\%} & \textbf{4.86\%} & \textbf{1.27\%} & \textbf{4.09\%} & \textbf{1.06\%} & \textbf{3.34\%} \\[0.1em]
        & \textbf{Imp. (DiffRec)} & \textbf{4.37\%} & \textbf{2.64\%} & \textbf{4.96\%} & \textbf{3.00\%} & \textbf{23.06\%} & \textbf{22.70\%} & \textbf{22.31\%} & \textbf{19.50\%} & \textbf{14.71\%} & \textbf{10.76\%} & \textbf{10.27\%} & \textbf{12.57\%} & \textbf{9.47\%} & \textbf{8.58\%} & \textbf{16.09\%} & \textbf{15.02\%} & \textbf{9.69\%} & \textbf{9.66\%} & \textbf{10.80\%} & \textbf{10.61\%}\\
        \cmidrule{1-22}  
    \end{tabular*}
    \vspace{-3mm}
    \label{table:overall}
\end{table*}

\section{Experiments}
\subsection{Experimental Settings}
\begin{table}[h]\small
    \centering
    \vspace{-4mm}
    \tabcolsep=0.1cm
    \caption{Statistics of experimental data.}
    \vspace{-2mm}
    \begin{tabular*}{\columnwidth}{@{\extracolsep{\fill}}c|cccc}
    \hline
    \textbf{Dataset} &\textbf{\#User}  &\textbf{\#Item} &\textbf{\#Interaction} &\textbf{\%Density}\\ \hline\hline
    \textbf{LastFM}          &1,892  &17,632 &92,834     &0.2783   \\
    \textbf{Amazon-Beauty}   &22,364 &12,102 &198,502    &0.0733   \\
    \textbf{Douban-Book}     &13,024  &22,347 &792,062    &0.2721   \\
    \textbf{Gowalla}        &29,858 &40,981 &1,027,370  &0.0840   \\
    \textbf{Yelp2018}        &31,668 &38,048 &1,561,406  &0.1296   \\
    \hline
    \end{tabular*}
    \label{table:data}
    \vspace{-2mm}
\end{table}

\noindent \textbf{Datasets.} We use five real-world datasets to evaluate the performance of TV-Diff, including \textit{Lastfm}\footnote{https://github.com/librahu/HIN-Datasets-for-Recommendation-and-Network-Embedding\label{footnoot:HIN}}, \textit{Amazon-Beauty}\footnote{https://github.com/Coder-Yu/SELFRec\label{footnoot:selfrec}}, \textit{Douban-Book}\textsuperscript{\ref{footnoot:selfrec}}, \textit{Gowalla}\footnote{https://github.com/xiangwang1223/neural\_graph\_collaborative\_filtering} 
and \textit{Yelp2018}\textsuperscript{\ref{footnoot:selfrec}} for overall comparisons. These datasets are publicly available and have been widely used in various literature~\cite{LightGCN,SGL,LightGCL,XSimGCL,DiffRec,BSPM}. All ratings are binarized as implicit feedback. We randomly split the datasets into training sets and test sets with a proportion of 8:2. 

\noindent \textbf{Metrics.} We use two popular evaluation metrics for Top-K recommendations: \textit{Recall@K (R@K)} and \textit{NDCG@K (N@K)}, where $K\in \{10,20\}$. We highlight the best method in boldface and underline the follow-up one. In our work, a relative improvement above 1\% is considered significant~\cite{Significant}. 

\noindent \textbf{Baselines.}
We conduct experiments by comparing a series of representative recommender models (RMs), categorized into five types: \textit{Base RM} utilizes its ability to simply extract collaborative signals on the user-item interactions: \textbf{BPR-MF}~\cite{BPR}, \textbf{NeuMF}~\cite{NeuMF}; \textit{Autoencoder-based RM} learns compressed representations and reconstructs incomplete interactions by an encoder and a decoder respectively: \textbf{CDAE}~\cite{CDAE}, \textbf{MultiVAE}~\cite{MultiVAE}; \textit{Graph-based RM} mainly focus on the message-passing mechanism and their efficiency on graphs: \textbf{LightGCN}~\cite{LightGCN}, \textbf{ChebyCF}~\cite{ChebyCF}; \textbf{LinkProp}~\cite{LinkProp}, \textbf{SGCL}~\cite{SGCL}. \textit{Negative Sampling RM} aims to sample or synthesize informative negative items from the large item pool: \textbf{MixGCF}~\cite{MixGCF}, \textbf{AHNS}~\cite{AHNS}; \textit{Diffusion-based RM} leverages iterative corruption and denoising to model a refined user preference: \textbf{CODIGEM}~\cite{CODIGEM}, \textbf{DiffRec}~\cite{DiffRec}, \textbf{BSPM}~\cite{BSPM}, \textbf{GiffCF}~\cite{GiffCF}, \textbf{DDRM}~\cite{DDPM}, \textbf{HDRM}~\cite{HDRM}.

\noindent \textbf{Implementation Details.}
For a fair comparison, we use grid search to fine-tune all the hyperparameters of the baselines within the intervals reported in the original papers, and assign the same threshold of early-stopping (i.e., 10~\cite{DiffRec}) for all models. For the general settings, we fix a series of hyperparameters for all comparison methods: the latent size is set to 64, the coefficient of regularization is fixed to $1e{-4}$, the relaxation rate is empirically set to 3~\cite{gumbel_softmax}, and the number of inference timesteps is equal to the training timesteps~\cite{DiffRec}. We use the Xavier method~\cite{Xavier} to initialize all parameters, and Adam~\cite{Adam} to optimize all these models. For fair negative sampling evaluations, we uniformly pair a positive item with one negative item for each training iteration~\cite{BPR}. To reproduce the results, we fix the random seed to 0 in PyTorch. For more detailed information, please refer to the source code~\footnote{https://github.com/umsimonchen/TV-Diff}. 

\subsection{Performance Comparison}
In this section, we compare our TV-Diff with the five different kinds of recommender models. Improvements compared to the follow-up models and the backbone DiffRec are also highlighted for inference. According to the experimental results, we can obtain the following observations:

$\bullet$ TV-Diff consistently yields the best performance on all datasets in terms of all evaluation metrics (cf. Table~\ref{table:overall}). This demonstrates that TV-Diff is capable of distilling, modeling, and generating comprehensive user preferences. We advocate that TV-Diff bestows a holistic diffusion recommendation framework via (1) a complete optimization objective, (2) a bipartite-graph-specific denoiser, and (3) an adaptive DM-based hard negative sampling strategy. 

$\bullet$ DM-based recommender models exhibit superior performance over autoencoder-based models, attributable to their distinctive denoising score matching mechanism. Specifically, TV-Diff achieves significant improvements, averaging over 12\%, compared to existing DM-based approaches. These gains demonstrate the ability of TV-Diff to overcome inherent limitations from the thermodynamic and topological views, which hinder diffusion models from attaining optimal results. Furthermore, by eliminating the need for pre-training, TV-Diff mitigates the instability that is highly sensitively associated with the quality of pre-trained embeddings.

$\bullet$ We acknowledge the admirable performance of the contrastive learning approach, which consistently achieves competitive results as a follow-up model. However, its capability is widely attributed to batch-wise negative sampling, leveraging multiple negatives to capture comprehensive semantics. Notably, TV-Diff outperforms the contrastive learning and negative sampling approaches with an average improvement exceeding 3.5\%. This result suggests that our proposed hard negative sampling strategy effectively identifies the most informative negative while preventing redundant batch-wise negative sampling and their computations. 

\begin{figure}[!t]
  \centering
  \includegraphics[width=1.\columnwidth]{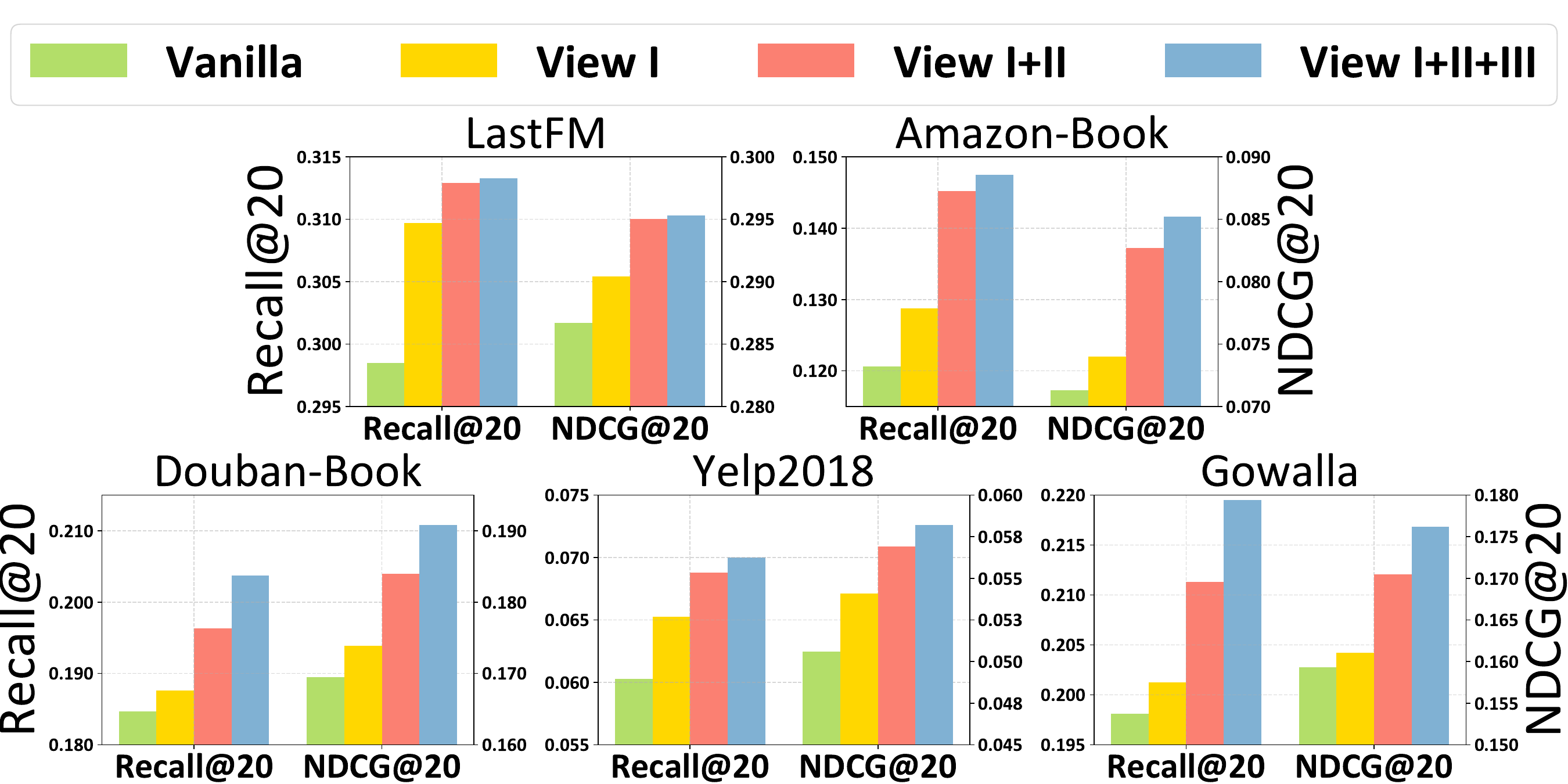}
  \vspace{-6mm}
  \caption{Ablation study on different views.}
  \vspace{-5mm}
  \label{fig:view}
\end{figure}

\subsection{Ablation Analysis}
\label{subsection:ablation}
\noindent \textbf{Effectiveness of Different Views.} 
We first analyze the interplay and contributions of TV-Diff’s components. Specifically, View I, II, and III denote the \textit{Helmholtz Free Energy} maximization, anisotropic denoiser, and acceptance-rejection Gumbel sampling process (AR-GSP), respectively. The ablated models are evaluated incrementally, corresponding to the framework's progressive development: \textit{Helmholtz Free Energy} maximization serves as the core foundation; the \textit{entropy}-based objective subsequently enlightens the anisotropic denoiser; and finally, the training process of the denoiser is intertwined with AR-GSP in terms of the reconstruction and semantics. Vanilla refers to the backbone model DiffRec, without any proposed view, and View I+II+III represents our complete framework.

From the results, we observe that incorporating newly different components based on the previous ablated models leads to significant performance improvements across all base models (cf. Table~\ref{fig:view}). This confirms that each component is crucial to TV-Diff. On one hand, the anisotropic denoiser achieves the highest gains among most datasets. For example, View I+II outperforms its counterpart View I by over 8\% on average. On the other hand, AR-GSP achieves only over 3\% improvements, and even less on smaller datasets (e.g., LastFM). These findings provide compelling evidence that the existing DM-based recommender models are markedly ineffective in capturing anisotropy among bipartite graphs, while their outstanding denoising ability somehow diminishes the need for hard negative sampling, thereby saving computational resources.

\begin{figure}[!h]
  \centering
  \vspace{-4mm}
  \includegraphics[width=1.\columnwidth]{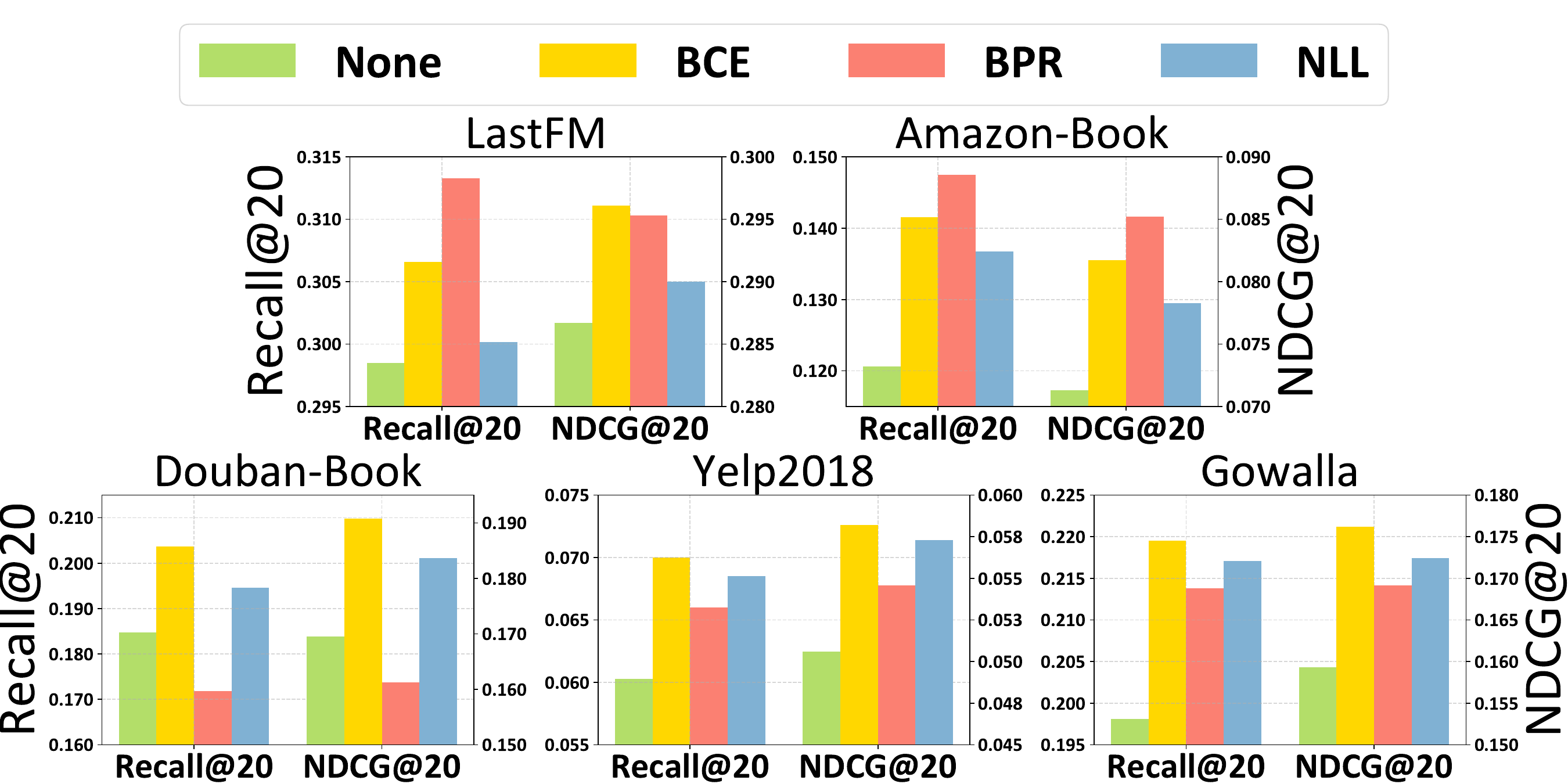}
  \vspace{-6mm}
  \caption{Ablation study on \textit{entropy}-based objectives.}
  \vspace{-4mm}
  \label{fig:entropy}
\end{figure}

\noindent \textbf{Impact of \textit{Entropy}-Based Objectives.}
Three widely used \textit{entropy}-based training objectives (i.e., BCE, BPR, NLL) are evaluated against a control group (i.e., None). The results (cf. Figure~\ref{fig:entropy}) show that \textit{entropy}-based objectives consistently enhance the performance compared to solely employing \textit{energy}-based objectives. Furthermore, our observations indicate that small datasets favor BPR, whereas large datasets favor BCE. We hypothesize that the triplet formulation in BPR provides more nuanced user preferences from limited interactions, but may induce overfitting and deteriorate generalization if given abundant interactions. 

\begin{table}[!h]
    \centering
    \renewcommand{\arraystretch}{0.9}
    \vspace{-2mm}
    \caption{Ablation study on Acceptance-Rejection Sampling (AR) compared with Sub-Linear Correlation (SL).}
    \vspace{-2mm}
    \begin{tabular*}{\columnwidth}{@{\extracolsep{\fill}}cc|cccc}
        \hline
        \textbf{Dataset} &\textbf{Method} & \textbf{R@10}    & \textbf{N@10}    &\textbf{R@20}    &\textbf{N@20}   \\
        \hline\hline
        \multirow{2}{*}{LastFM}
        & w/ SL & 0.2098 & 0.2419 & 0.3022 & 0.2895\\
        & w/ AR & \textbf{0.2165} & \textbf{0.2533} & \textbf{0.3108} & \textbf{0.2925}\\
        \hline
        \multirow{2}{*}{Douban}
        & w/ SL & 0.1274 & 0.1431 & 0.1768 & 0.1545\\
        & w/ AR  & \textbf{0.1353} & \textbf{0.1587} & \textbf{0.1832} & \textbf{0.1692}\\
        \hline
        \multirow{2}{*}{Gowalla}
        & w/ SL & 0.1318 & 0.1375 & 0.2072 & 0.1627\\
        & w/ AR  & \textbf{0.1437} & \textbf{0.1459} & \textbf{0.2178} & \textbf{0.1721}\\
        \hline
    \end{tabular*}
    \vspace{-2mm}
    \label{table:AR}
\end{table}

\noindent \textbf{Comparison of Acceptance-Rejection sampling and Traditional Hard NS.}
To rigorously evaluate the superiority of the acceptance-rejection sampling (AR) over the conventional sub-linear correlation (SL) in recommendation tasks, we conduct a direct comparison under identical conditions, where both sample one negative item across the entire item sets. For clarity in causal analysis, we employ LightGCN as the base model due to its concise architecture. Our results (cf. Table~\ref{table:AR}) state that AR consistently and significantly outperforms SL, where the improvements notably exceed even 10\% on Douban-Book. These findings highlight AR's superior efficacy in learning latent semantics within recommendation tasks and underscore the importance of truncating true negatives during sampling, particularly for long-tail distributions.

\begin{figure}[ht]
  \centering
  % \vspace{-4mm}
  \includegraphics[width=1.\columnwidth]{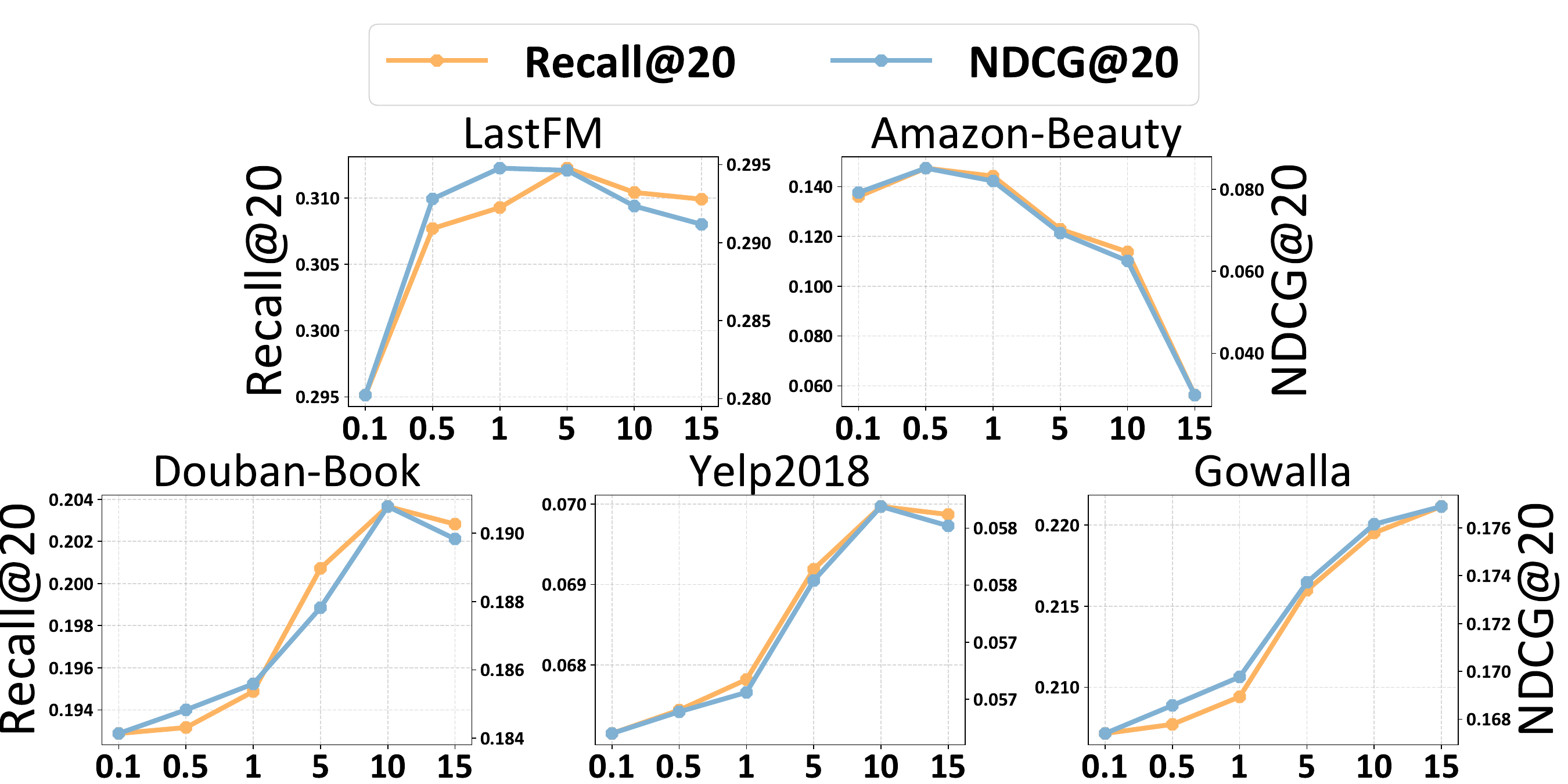}
  \vspace{-6mm}
  \caption{Influence of the temperature $t$.}
  \vspace{-3mm}
  \label{fig:temp}
\end{figure}

\noindent \textbf{Hyperparameter Sensitivity Analysis for Temperature $t$.}  
We analyze the sensitivity of the key hyperparameter: temperature $t$. This parameter controls the priority between \textit{energy} and \textit{entropy} for the maximization of \textit{Helmholtz free energy}. We vary $t$ in the range \{0.1, 0.5, 1, 5, 10, 15\}. According to the results (cf. Figure~\ref{fig:temp}), TV-Diff reaches a peak when $t=1$ for small datasets and $t=10$ for large datasets. This phenomenon suggests balanced optimization strategies are more preferable for small datasets, while greater importance should be assigned to \textit{entropy}-based objectives for large datasets, given the inherent sparsity per user/item of datasets.

\begin{figure}[h]
  \centering
  \vspace{-2mm}
  \includegraphics[width=1.\columnwidth]{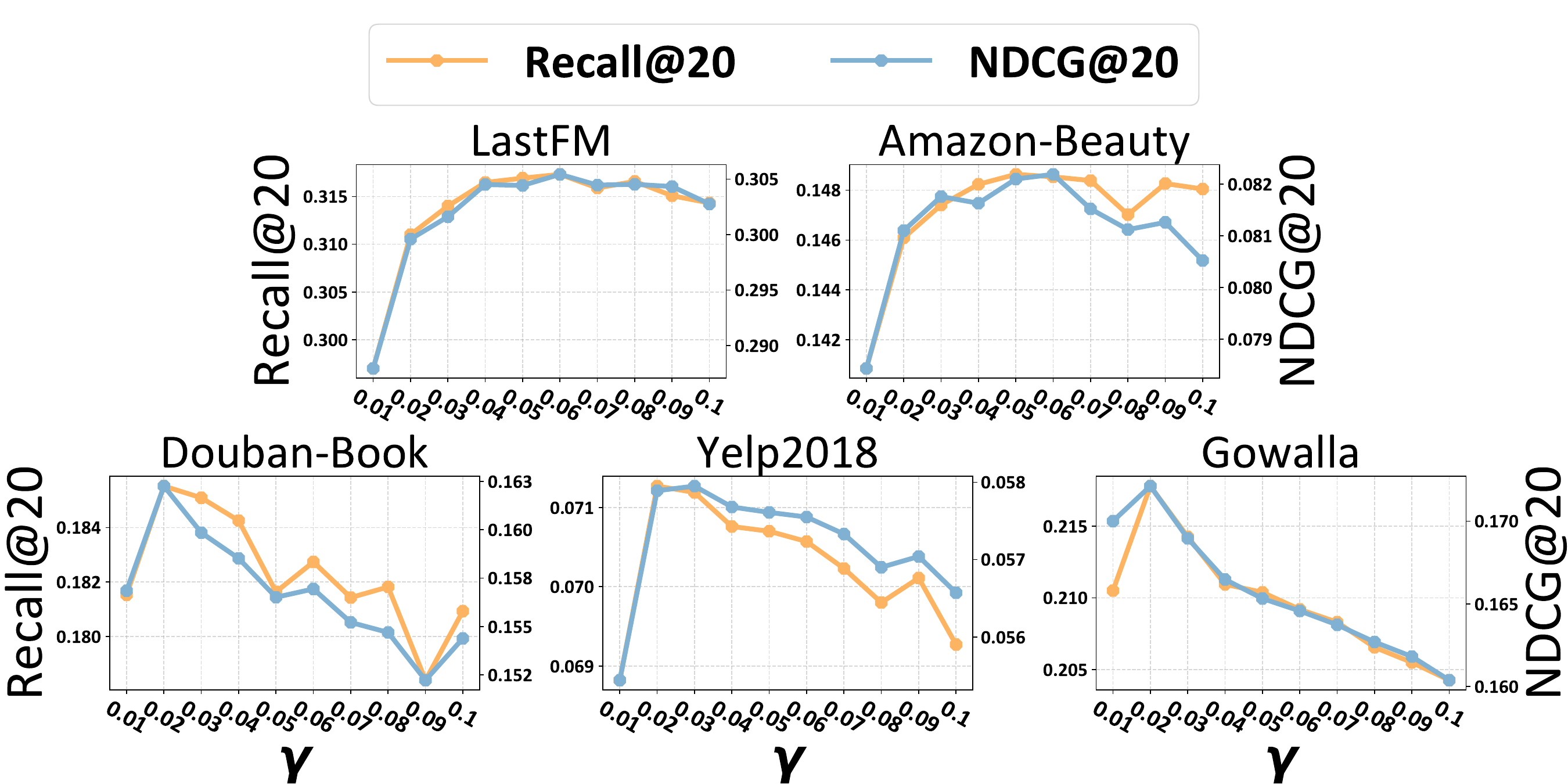}
  \vspace{-6mm}
  \caption{Influence of the negative factor $\gamma$.}
  \vspace{-3mm}
  \label{fig:gamma}
\end{figure}

\noindent \textbf{Hyperparameter Sensitivity Analysis for Negative Factor $\gamma$.}  
We also analyze the sensitivity of the negative factor $\gamma$. This parameter controls both the number and position of candidate negative items within the entire itemset. Hyperparameter $\gamma$ is ranged in [0.01, 0.1] in increments of 0.01. The findings indicate that TV-Diff’s performance consistently improves across all datasets as $\gamma$ increases, reaches a peak, and then gradually declines (cf. Figure~\ref{fig:gamma}). On one side, small datasets benefit from $\gamma=0.5$ to avoid representation collapse. On the other side, large datasets achieve the best performance when $\gamma=0.2$, reflecting their need for few but informative negatives following a long-tail distribution. It is worth noting that, when $\gamma$ becomes too small, performance declines sharply, suggesting items with higher recommendation scores may inadvertently include unobserved positive samples. Thus, to mitigate biased selection based on recommendation scores, this observation underscores the rationale for applying uniform sampling on candidate negative items after the completion of AR, as in our implementation.

\noindent \textbf{Case Study for Visualization of \textit{Energy} and \textit{Entropy}.}
Following the pilot experiments, we visualize the reconstructed interaction probability in terms of \textit{energy} and \textit{entropy}. According to Figure~\ref{fig:visual} and \ref{fig:delta}, we recognize that, as a DM-based recommendation framework, TV-Diff aligns with mainstream approaches in prioritizing \textit{energy} maximization. Nonetheless, TV-Diff exhibits a superior capability to minimize the \textit{entropy} compared to DiffRec, i.e., $\frac{\Delta S_D}{\Delta S_T}>1$. These results illustrate that TV-Diff transcends the limitation of a unilateral training objective, reconciles both factors via \textit{free energy}, and attains enhanced generalization and performance.

\begin{table} [!ht] %\scriptsize
    \centering
    % \vspace{-3mm}
    \tabcolsep=0.1mm
    \caption{Efficiency comparison. \textbf{UT} denotes the Unit Time per epoch; \textbf{Ep.} denotes the training Epoch; \textbf{K} denotes layers; $|\textbf{\textit{R}}|^+$ denotes the non-zero values of the graph.} 
    \vspace{-2mm}
    \begin{tabular*}{\columnwidth}{@{\extracolsep{\fill}}c|cc|cc|cc} \hline
    &\multicolumn{2}{c}{\textbf{LightGCN}} &\multicolumn{2}{c}{\textbf{DiffRec}} &\multicolumn{2}{c}{\textbf{TV-Diff}}\\
    \cmidrule{2-3} \cmidrule{4-5} \cmidrule{6-7}
    % \cmidrule
    \textbf{(s=second)} &\textbf{UT}$\downarrow$ &\textbf{\#Ep.}$\downarrow$ &\textbf{UT}$\downarrow$ &\textbf{\#Ep.}$\downarrow$ &\textbf{UT}$\downarrow$ &\textbf{\#Ep.}$\downarrow$\\ \hline \hline
    LastFM & 0.623s & 424 & 0.353s & 46 & 1.056s & 26\\ \hline
    Douban & 5.345s & 270 & 2.572s & 66 & 1.649s & 42 \\ \hline
    Gowalla & 14.748s & 368 & 19.303s & 113 & 5.162s & 45 \\ \hline\hline
    Space &\multicolumn{2}{c}{$\mathcal{O}((m+n)d)$} &\multicolumn{2}{c}{$\mathcal{O}((m+n)d)$} &\multicolumn{2}{c}{$\mathcal{O}((m+n)d)$}\\ \hline
    Time &\multicolumn{2}{c}{$\mathcal{O}(|\textbf{\textit{R}}^{+}|Kd)$} &\multicolumn{2}{c}{$\mathcal{O}(2mnd)$} &\multicolumn{2}{c}{$\mathcal{O}(|\mathcal{B}|nd+|\textbf{\textit{R}}^{+}|d)$}\\ \hline
    \end{tabular*}
    \vspace{-3mm}
    \label{table:complexity}
\end{table}

\noindent \textbf{Efficiency Analysis.} For evaluations, LightGCN and DiffRec are chosen due to their simplest architecture. We run all on an Intel(R) Core(TM) i7-12700 CPU and a GeForce RTX 3090 GPU. According to the analysis (cf. Table~\ref{table:complexity}), TV-Diff is fairly efficient, especially on large datasets where LightGCN suffers from multi-layer message-passing on large graphs and DiffRec fails to maintain performance from batch-training. Notably, the ranking for AR-GSP occupied $\mathcal{O}(|\mathcal{B}|n\log n)$, which can be omitted. 

\section{Related Work}
\subsection{Diffusion-based Recommender Models}
Diffusion models are viewed as extensions of flow models~\cite{flow_model}, autoencoders~\cite{AE}, and variational autoencoders~\cite{VAE}, architecturally evolving single discrete encoder/decoder steps into multiple continuous diffusion steps. From an optimization perspective, they also draw upon deep energy-based models~\cite{EBM}, score-based models (SBMs)~\cite{SBM}, and denoising SBMs~\cite{denoisingSBM} to maximize the likelihood in a more tractable and efficient manner. Inspired by the breakthroughs of~\cite{thermoDiff,DDPM,SDE,MultiDiff}, diffusion models have recently been applied to general recommendation tasks~\cite{CODIGEM,DiffRec} and demonstrated their outstanding potential. However, early diffusion-based recommender models overlook the topology information inherent in bipartite interaction graphs, a consideration largely irrelevant in the field of computer vision. To address this, BSPM~\cite{BSPM}, GiffCF~\cite{GiffCF}, and S-Diff~\cite{S-Diff} simulate the message-passing within the diffusion process. Additionally, a two-stage training strategy~\cite{DDRM,HDRM} also facilitates graph structures learning during pre-training and high-quality representation retrieval via diffusion-based fine-tuning. Nevertheless, this approach incurs substantially higher computational costs than end-to-end training.

\vspace{-1mm}
\subsection{Graph-based Recommender Models}
With the rapid development of graph neural networks (GNNs)~\cite{GCN,GAT,GraphSAGE}, graph-based recommendation models have become thriving~\cite{PinSAGE,LightGCN}. Generally, four mainstream approach prevail: (1) Graph Signal Processing~\cite{PGSP,SGFCF,ChebyCF} methods leveage Graph Fourier Transform and low-pass filters to extract low-frequency collaborative signals embedded in graph structures; (2) Hypergraph methods~\cite{DHCF,HCCF,EIISRS} construct hypergraphs to explore richer collaborative filtering architectures capable of modeling higher-order ralationships; (3) Negative Sampling methods~\cite{MixGCF,DENS,AHNS} focus on identifying the underlying negative distribution and maximizing the distiction between positive and negative items by exploiting graph structures; (4) Contrastive Learning models~\cite{SGL,LightGCL,SGCL} harness alignment and uniformity of representations across augmented graph views to mitigate the data sparsity. However, these methods rely on the multi-layer message-passing for noise smoothing and feature extraction, bearing substantial overhead on massive graphs. 

\vspace{-1mm}
\subsection{Hard Negative Sampling}
As a necessary component in recommender systems, negative sampling critically impacts recommendation outcomes~\cite{NSSurvey}. The most common approach, Random Negative Sampling (RNS)~\cite{BPR}, often involves irrelevant items during training. This limitation has motivated the development of Hard Negative Sampling (Hard NS), which prioritizes informative negative items. Theoretical analyses for link prediction tasks indicate that hard negative distributions should exhibit sub-linear correlation with positive distributions~\cite{MCNS,MixGCF,SRNS,AHNS,DMNS}. However, applying this principle to BPR can be problematic: it risks deteriorating the pairwise loss into a pointwise objective, potentially leading to worse performance. In this work, we pioneer the exploration of hard negative sampling in the context of diffusion-based recommendation tasks. 

\section{Conclusion}
In this work, we introduce \textbf{TV-Diff}, a novel tri-view diffusion recommendation framework designed to address the limitations of DM-based recommender models. By leveraging the \textit{Helmholtz free energy maximization}, TV-Diff optimizes recommender models by both \textit{energy} and \textit{entropy}, where the robustness of the joint-training is guaranteed by the anisotropic denoiser and acceptance-rejection Gumbel sampling process. 
The former captures nuances between user and item degree distributions and reconstructs predictions by user-item cross-correlation with explicit topology information, while the latter provides informative negatives for DM-based recommender models. 
In the future, we will delve into the fundamental architecture and process of DMs for better recommendations.

\begin{acks}
This work was supported by The Science and Technology Development Fund, Macau SAR (0126/2024/RIA2, 001/2024/SKL), CG-2026, GDST (2020B1212030003, 2023A0505030013), Nansha District (2023ZD001), MYRG-GRG2023-00186-FST-UMDF, MYRG-GRG2025-00234-FST.
\end{acks}

\bibliographystyle{ACM-Reference-Format}
\bibliography{sample-base}

\appendix
\section{Discussion on Helmholtz Free Energy}
\begin{table*}[!ht]
    \centering
    \caption{Comparison between thermodynamics and informatics in terms of Helmholtz free energy.}
    \vspace{-3mm}
    \begin{tabular*}{\textwidth}{@{\extracolsep{\fill}}c|c|c|c}
        \hline
        \multicolumn{4}{c}{\textbf{\textit{Helmholtz Free Energy}}\qquad $\textbf{H}\uparrow \equiv \textbf{U}\uparrow - \textbf{t}\textbf{S}\downarrow$}\\ \hline
        \textbf{Acronym} & \textbf{Thermodynamics} & \textbf{Informatics} & \textbf{Target}\\
        \hline\hline
        \textbf{H} & \textit{Free energy to be obtained for useful work} & \textit{Generalized useful pattern for accurate prediction} & \textit{Performance} \\
        \textbf{U} & \textit{Internal energy of a thermodynamic system} & \textit{Information quantity of the training data} & \textit{Reconstruction} \\
        \textbf{S} & \textit{Entropy causing spontaneous processes} & \textit{Magnitude of uncertainty in information} & \textit{Refinement} \\
        \textbf{t} & \textit{Given temperature of the system} & \textit{Given priorities of two metrics in model} & \textit{Coefficient} \\
        \hline
    \end{tabular*}
    \vspace{-3mm}
    \label{table:free_energy}
\end{table*}
\textit{Helmholtz free energy} is originally defined via a \textit{Legendre transformation}~\cite{legendre} on the fundamental thermodynamic relation. Starting from the internal \textit{energy} \textbf{U}, a function of \textit{entropy} \textbf{S} and \textit{volume} \textbf{V}, the transformation replaces \textit{energy} and \textit{entropy} with their conjugate variable, temperature \textbf{t} $=(\partial U/\partial S)|_{V}$, as the independent variable for volume-agnostic processes. Over the years, it has been adopted across a variety of domains, such as chemistry~\cite{chemistry}, biology~\cite{biology}, and computer graphics~\cite{graphics}. \citeauthor{Free-Energy}~\cite{Free-Energy} first introduced this concept into machine learning. They illustrated that the objective function minimized by autoencoders is mathematically equivalent to optimizing a bound on the description length under the Minimum Description Length (MDL) principle, minimizing the \textit{Helmholtz free energy} statistically in the meanwhile. Thus, the MDL goal and the variational approximation of the \textit{Helmholtz free energy} manifest as a trade-off optimized by autoencoders, where the reconstruction accuracy term is balanced by a constraint on the capacity of the latent representation. In spite of previous work, our work mainly focuses on diffusion- and graph-based recommendation tasks, inspired by a series of our designated experiments. For better understanding, we summarize the similarities in \textit{Helmholtz free energy} between thermodynamics and informatics (cf. Table~\ref {table:free_energy}).

\section{Proof for Equivalence of Entropy and BPR}
\label{proof:BPR}
As \textit{entropy}-based training objectives, Binary Cross Entropy (BCE) explicitly optimizes the models by decreasing the \textit{entropy}, while the mechanism of BPR is subtle. According to Observation 2 (cf. Figure~\ref {fig:visual} \&~\ref{fig:delta}), we experimentally prove the equivalence between \textit{entropy} and BPR. Here, we derive it theoretically:

\begin{theorem}
    \vspace{-0.5\baselineskip}
    BPR loss is equivalent to the \textit{entropy}, that is, model optimization by BPR loss will reduce the entropy in the meantime.
\end{theorem}
\vspace{-0.5\baselineskip}

\begin{proof}
    \vspace{-0.5\baselineskip}
    By definition, we have:
    \begin{equation}
        S(u)=-\hat{\textbf{\textit{r}}}_u\log\hat{\textbf{\textit{r}}}_u^{\top},
    \end{equation}
    where $\hat{\textbf{\textit{r}}}_u=\exp({\tilde{\textbf{\textit{r}}}}_u)/\Vert \exp({\tilde{\textbf{\textit{r}}}}_u)\Vert_1=\exp({\tilde{\textbf{\textit{r}}}}_u)/(n\cdot \mathbb{E}[\exp({\tilde{r}}_{u,*})])$. Since the entropy computation is based on probability vectors $\hat{\textbf{\textit{r}}}_u$, we reformulate the BPR loss in the form of vectors by $\tilde{{\textbf{\textit{r}}}}_u$ as well:
    \begin{equation}
        \begin{split}
            \mathcal{L}_{BPR}(u)&= \sum_{i\in N(u),j\notin N(u)} -\log\sigma(\tilde{r}_{u,i}-\tilde{r}_{u,j}),\\
            &\equiv \Vert- \log\sigma({\tilde{\textbf{\textit{r}}}}_u-\mathbb{E}[\tilde{r}_{u,*}])\Vert_1,
        \end{split}
    \end{equation}
    where we substitute the score of negative items with the expectation of score among all items $\mathbb{E}[\tilde{r}_{u,*}]$ for user $u$. The rationales for it are: (1) item sets usually adhere to the long-tail distribution during to the high sparity, where most items are non-interactive, i.e., $\mathbb{E}[\tilde{r}_{u,-}]\approx\mathbb{E}[\tilde{r}_{u,*}]$; and (2) the score of interactive items is always non-negative during training, i.e., $\tilde{r}_{u,+}\geq0$, making the cost for mistaking the interactive items for negative samples bounded. Hence, this substitution is harmless to the model optimization. 
    
    When BPR loss converges:
    \begin{equation}
        \frac{\partial\mathcal{L}_{BPR}}{\partial{\tilde{\textbf{\textit{r}}}}_u}=-\frac{1}{\sigma(\cdot)} \cdot \frac{\partial\sigma(\cdot)}{\partial{\tilde{\textbf{\textit{r}}}}_u}=-\frac{1}{\sigma(\cdot)}\cdot(1-\sigma(\cdot))=-\frac{1}{\sigma(\cdot)}+ 1=\vmathbb{0} ,
    \end{equation}
    \begin{equation}
    \begin{split}
        \sigma({\tilde{\textbf{\textit{r}}}}_u-\mathbb{E}[\tilde{r}_{u,*}])&=\vmathbb{1}\\
        \exp(-({\tilde{\textbf{\textit{r}}}}_u-\mathbb{E}[\tilde{r}_{u,*}]))&=\vmathbb{0}\\
        \exp(-{\tilde{\textbf{\textit{r}}}}_u)\cdot\exp(\mathbb{E}[\tilde{r}_{u,*}])&=\vmathbb{0}.
    \end{split}
    \end{equation}
    According to the \textit{Jensen's Inequality}, we have:
    \begin{equation}
        \exp(\mathbb{E}[\tilde{r}_{u,*}])\leq\mathbb{E}[\exp(\tilde{r}_{u,*})],
    \end{equation}
    since $\exp(\cdot)$ is a convex function. Hence, we decompose the BPR convergence condition into two sub-conditions: \textcircled{1} Equality holds for \textit{Jensen's Inequality}, i.e, $\tilde{r}_{u,1}=\cdots=\tilde{r}_{u,n}$, and \textcircled{2} gradient of the updated BPR loss becomes 0, i.e., $\exp(-{\tilde{\textbf{\textit{r}}}}_u)\cdot \mathbb{E}[\exp(\tilde{r}_{u,*})] =\vmathbb{0}$.
    Thus, the definition of \textit{entropy} can be updated as follows:
    \begin{equation}
    \begin{split}
        S(u)&=-\frac{\exp({\tilde{\textbf{\textit{r}}}}_u)}{n\cdot \mathbb{E}[\exp(\tilde{r}_{u,*})]}(\log\frac{\exp(\tilde{\textbf{\textit{r}}}_u)}{n\cdot \mathbb{E}[\exp(\tilde{r}_{u,*})]})^{\top}\\
        =\frac{1}{n}&\cdot \underbrace{\frac{\exp(\tilde{\textbf{\textit{r}}}_u)}{\mathbb{E}[\exp(\tilde{r}_{u,*})]}}_{\color{orange} \to \vmathbb{1} \quad \mbox{for \textcircled{1}}} (\log\underbrace{(\frac{\mathbb{E}[\exp(\tilde{r}_{u,*})]}{\exp(\tilde{\textbf{\textit{r}}}_u)})}_{\color{orange}\to \vmathbb{0} \quad \mbox{for \textcircled{2}}}+\underbrace{\log n}_{\color{orange}\mbox{const.}})^{\top} \to -\infty,
        \label{eq:bpr}
    \end{split}
    \end{equation}
    where $\lim_{x\to0}\log x=-\infty$.
    \pushQED{}
\end{proof}
\vspace{-0.5\baselineskip}

\begin{remark}
    \vspace{-0.5\baselineskip}
    Regarding a stable training process, the first term (cf. Eq. (\ref{eq:bpr})) cannot satisfy condition $\textcircled{2}$ as it will inevitably confront a gradient explosion when 0 serves as the denominator of fractions. Besides, to hold both conditions, the condition $\textcircled{2}$ has to be assigned to the second term as a necessity. Although the co-constraints of both conditions preclude a closed-form solution, the BPR loss exhibits monotonic optimization behavior alongside decreasing \textit{entropy} in this configuration.
\end{remark}

\section{Proof for Higher Entropy on Multi-Layer Message-Passing}
\label{proof:MP}
To explore the higher \textit{entropy} achieved via multi-layer message-passing, we first rigorously characterize its higher smoothness compared with single-layer message-passing. It is worth noting that the message-passing mechanism we are about to prove is a linear process as defined by LightGCN~\cite{LightGCN}.

\begin{proposition}[~\cite{graph_spectral_reference}]
    \vspace{-0.3\baselineskip}
    Given a graph $\mathcal{G}$ with its corresponding normalized Laplacian matrix $\bar{\textbf{\textit{L}}}$, the eigenvalues of $\bar{\textbf{\textit{L}}}$ are $\bar{\lambda}_i \in [0,2]$, such that the eigenvalues of the symmetric normalized adjacency matrix $\bar{\textbf{\textit{A}}}= \begin{bmatrix}\vmathbb{0}&\bar{\textbf{\textit{R}}}\\\bar{\textbf{\textit{R}}}^{\top}&\vmathbb{0}\end{bmatrix} \in \mathbb{R}^{(m+n)\times (m+n)}$ after eigendecomposition are: $-1 \leq (1-\bar{\lambda}_1) \leq \cdots \leq (1-\bar{\lambda}_{m+n}) \leq 1.$
\end{proposition}
\vspace{-0.5\baselineskip}
    
\begin{proposition}
    \label{lemma:over-smoothing}
    \vspace{-0.3\baselineskip}
    When the over-smoothing phenomenon in BP-based graph learning occurs, all low and middle frequencies of signals, except for the lowest and highest ones, will be removed, i.e., $\forall\bar{\lambda}_i \in (0,2), (1-\bar{\lambda}_i)^{\infty} \to 0$.
\end{proposition}
\vspace{-0.3\baselineskip}

\begin{theorem}
    \vspace{-0.3\baselineskip}
    \label{theorem:frequency}
    \textit{Single-layer message-passing learns NO less different frequencies of signals than multi-layer message-passing, i.e.,} $(\forall \bar{\lambda}_i \in [0, 1]) \land (\forall \alpha_k>0) \land (\forall K \in \mathbb{Z}^+), (1-\bar{\lambda}_i) \geq [\sum^{K}_{k=1}\alpha_k(1-\bar{\lambda}_i)^k]/\sum^K_{k=1}{\alpha_k}$, where $\alpha_k$ denotes the attention score of layer $k$.
\end{theorem}
\vspace{-0.5\baselineskip}

\begin{proof}
    \vspace{-0.5\baselineskip}
    We prove it by mathematical induction.\\ 
    First, let $\mathcal{F}^{'}_G(K)=[\sum^{K}_{k=1}\alpha_k(1-\bar{\lambda}_i)^k]/\sum^K_{k=1}{\alpha_k}$, $\mathcal{F}_{G}=1-\tilde{\lambda}_i$.\\
    When $K=1$, $\mathcal{F}^{'}_{G}(1) = (1-\bar{\lambda}_i)$. Hence, $\mathcal{F}^{'}_{G}(1) \leq \mathcal{F}_{G}$ is true.\\
    Assume $\mathcal{F}^{'}_G(K) \leq \mathcal{F}_G$ for all $K > 1$. Then, we can prove that:
    \begin{align}
        \mathcal{F}^{'}_G(K+1) = [\sum^{K}_{k=1}\alpha_k&(1-\bar{\lambda}_i)^k + \alpha_{K+1}(1-\bar{\lambda_i})^{K+1}]/[\sum^{K}_{k=1}{\alpha_k}+\alpha_{K+1}], \nonumber\\[-2mm]
        \mathcal{F}_G - \mathcal{F}^{'}_G(K+1) &=[\sum^{K}_{k=1}{\alpha_k}(1-\bar{\lambda_i}) - \sum^{K}_{k=1}\alpha_k(1-\bar{\lambda}_i)^k] \label{eq:frequency1}\\
        + [\alpha_{K+1}(1-&\bar{\lambda_i}) - \alpha_{K+1}(1-\bar{\lambda_i})^{K+1}] \geq 0 . \label{eq:frequency2}
    \end{align}
    With assumptions, the first term (cf. Eq. (\ref{eq:frequency1})) and the second term (cf. Eq. (\ref{eq:frequency2})) are greater than or equal to 0 under the given Proposition~\ref{lemma:over-smoothing}. Hence, $\mathcal{F}_G \geq \mathcal{F}^{'}_G(K)$ for all $\bar{\lambda}_i \in [0,1]$ when $K \in \mathbb{Z}^{+}$ and $\alpha_k > 0$. Similar conclusions can be drawn for high-frequency signals, i.e., $\bar{\lambda}_i \in [1,2]$, since we only focus the magnitude ($|\mathcal{F}_G| \geq |\mathcal{F}^{'}_G(K)|$) rather than the direction (similar or dissimilar).
    \pushQED{}
\end{proof}
\vspace{-0.3\baselineskip}

\begin{remark}
    \vspace{-0.5\baselineskip}
    We omit the input embedding since we empirically find significant performance gains after skipping it in practice. Due to the absence of an associated filter, the input embedding leads to unstable training. Nonetheless, the proof with input embedding is still derivable, where both methods shall share the same scores $\alpha_0$ and $\alpha_1$.
\end{remark}
\vspace{-0.3\baselineskip}

\begin{proposition}
    \vspace{-0.3\baselineskip}
    Given the graph quadratic form $\mathcal{S}(\textbf{\textit{x}})=\textbf{\textit{x}}^{\top}\bar{\textbf{\textit{L}}}\textbf{\textit{x}}$ where $\bar{\textbf{\textit{L}}}$ is the Laplacian matrix, a lower value of $\mathcal{S}(\textbf{\textit{x}})$ indicates a smoother signal $\textbf{\textit{x}}$~\cite{quadratic_form}. Thus, $\nabla\mathcal{S}(\textbf{\textit{x}}\mid\textbf{\textit{x}}{'})=\mathcal{S}(\textbf{\textit{x}})\mathcal{S}^{-1}(\textbf{\textit{x}}{'})$ represents the difference of smoothness between $\textbf{\textit{x}}$ and $\textbf{\textit{x}}{'}$: $\nabla\mathcal{S}(\textbf{\textit{x}}\mid\textbf{\textit{x}}{'}) > 1$ states signal $\textbf{\textit{x}}$ is sharper than signal $\textbf{\textit{x}}{'}$, and vice versa.
\end{proposition}
\vspace{-0.3\baselineskip}

\begin{theorem}
    \vspace{-0.3\baselineskip}
    Single-layer message-passing fosters a NO smoother reconstructed joint scores matrix $\tilde{\textbf{\textit{A}}}_{S}$ than $\tilde{\textbf{\textit{A}}}_{M}$ that is retrieved from in multi-layer message-passing, i.e., $\nabla\mathcal{S}(\tilde{\textbf{\textit{A}}}_{S}\mid\tilde{\textbf{\textit{A}}}_{M}) \geq \textbf{I}$.
\end{theorem}
\vspace{-0.3\baselineskip}

\begin{proof}
    \vspace{-0.3\baselineskip}
    First, since the logarithm and sigmoid are element-wise and monotone-increasing functions, we have the definition of raw recommendation predictions:
    \begin{equation}
        \begin{split}
            &\tilde{a}_{u,i} =\log\sigma(\tilde{a}_{u,i}) \propto \textbf{\textit{e}}_u\textbf{\textit{e}}_i^{\top}\\
            \iff &\tilde{\textbf{\textit{A}}} = \log\sigma(\textbf{\textit{E}}_{final}\textbf{\textit{E}}_{final}^{\top}) \propto \textbf{\textit{E}}_{final}\textbf{\textit{E}}_{final}^{\top},
        \label{eq:smooth1}
        \end{split}
    \end{equation}
    where $\sigma(\cdot)$ denotes the sigmoid function. Hence, the pivotal point of proof lies in the similarity matrix derived from the self-multiplication of the final embedding $\textbf{\textit{E}}_{final}=\begin{bmatrix}
    \textbf{\textit{E}}_{\mathcal{U}}\\ \textbf{\textit{E}}_{\mathcal{I}}
    \end{bmatrix}\in \mathbb{R}^{(m+n)\times d}$. 

    \noindent Next, let $\bar{\textbf{\textit{A}}'}=
    [\sum^K_{k=1}\alpha_k\bar{\textbf{\textit{A}}}^{k}]/{\sum^K_{k=1}\alpha_k}$ as the equivalent adjacency matrix for multi-layer message-passing. Also, let $\textbf{\textit{X}} = \textbf{\textit{E}}_{final}\textbf{\textit{E}}_{final}^{\top}=\bar{\textbf{\textit{A}}}\textbf{\textit{E}}_0\textbf{\textit{E}}_0^{\top}\bar{\textbf{\textit{A}}}^{\top}$ and $\textbf{\textit{X}}{'} = \textbf{\textit{E}}'_{final}\textbf{\textit{E}}'^{\top}_{final}=\bar{\textbf{\textit{A}}{'}}\textbf{\textit{E}}_0\textbf{\textit{E}}_0^{\top}\bar{\textbf{\textit{A}}'}^{\top}$ denote the similarity matrices from single-layer and multi-layer message-passing, respectively. Hence, we have:
    \begin{equation}
        \begin{split}
            \textbf{\textit{X}}\textbf{\textit{X}}'^{-1} &= \bar{\textbf{\textit{A}}}\textbf{\textit{E}}_0\textbf{\textit{E}}_0^{\top}\bar{\textbf{\textit{A}}}^{\top}(\bar{\textbf{\textit{A}}'}\textbf{\textit{E}}_0\textbf{\textit{E}}_0^{\top}\bar{\textbf{\textit{A}}'}^{\top})^{-1} \\
            &\geq\bar{\textbf{\textit{A}}}\textbf{\textit{E}}_0\textbf{\textit{E}}_0^{\top} \textbf{\textit{I}} (\bar{\textbf{\textit{A}}'}\textbf{\textit{E}}_0\textbf{\textit{E}}_0^{\top})^{-1} = \bar{\textbf{\textit{A}}}(\bar{\textbf{\textit{A}}'}^{\top})^{-1} \geq \textbf{\textit{I}},
            \label{eq:smooth2}
        \end{split}
    \end{equation}
    where $\bar{\textbf{\textit{A}}}\bar{\textbf{\textit{A}}'}^{\top}=\bar{\textbf{\textit{U}}}(\textbf{\textit{I}}-\bar{\boldsymbol{\Lambda}})(\textbf{\textit{I}}-\bar{\boldsymbol{\Lambda}}')^{-1}\bar{\textbf{\textit{U}}}^{\top} \geq \bar{\textbf{\textit{U}}}(\textbf{\textit{I}}-\bar{\boldsymbol{\Lambda}}')(\textbf{\textit{I}}-\bar{\boldsymbol{\Lambda}}')^{-1}\bar{\textbf{\textit{U}}}^{\top} = \textbf{\textit{I}}$ since $(\textbf{\textit{I}}-\bar{\boldsymbol{\Lambda}}) \geq (\textbf{\textit{I}}-\bar{\boldsymbol{\Lambda}'}) $ by Theorem~\ref{theorem:frequency} and $\bar{\textbf{\textit{U}}}$ is the orthogonal matrix, i.e., $\bar{\textbf{\textit{U}}}^{\top}=\bar{\textbf{\textit{U}}}^{-1}$.

    \noindent Then, we delve into the difference of smoothness between $\textbf{\textit{X}}$ and $\textbf{\textit{X}}'$ by Eq. (\ref{eq:smooth2}), we have:
    \begin{equation}
        \begin{split}
            \nabla\mathcal{S}(\tilde{\textbf{\textit{A}}}_S\mid\tilde{\textbf{\textit{A}}}_{M})&=\textbf{\textit{X}}^{\top}\bar{\textbf{\textit{L}}}\textbf{\textit{X}}(\textbf{\textit{X}}'^{\top}\bar{\textbf{\textit{L}}}\textbf{\textit{X}}')^{-1} \\
            &\geq \textbf{\textit{X}}^{\top}\bar{\textbf{\textit{L}}}\textbf{\textit{I}}(\textbf{\textit{X}}'^{\top}\bar{\textbf{\textit{L}}})^{-1} = \textbf{\textit{X}}^{\top}(\textbf{\textit{X}}'^{\top})^{-1} \geq \textbf{\textit{I}}. 
            \label{eq:smooth3}
        \end{split}
    \end{equation} 
    By Eq. (\ref{eq:smooth1}) and Eq. (\ref{eq:smooth3}), we can see that the whole scores matrix in single-layer message-passing is NO smoother than those in multi-layer message-passing. With the increment of layers, the scores for all items tend to be similar and move towards a uniform distribution. Consequently, $entropy$ increases towards the maximum, potentially resulting in the over-smoothing issue~\cite{Over-Smoothing}.
    \pushQED{}
\end{proof}
\vspace{-0.3\baselineskip}

\begin{figure}[!ht]
  \centering
  \vspace{-4mm}
  \includegraphics[width=1.\columnwidth]{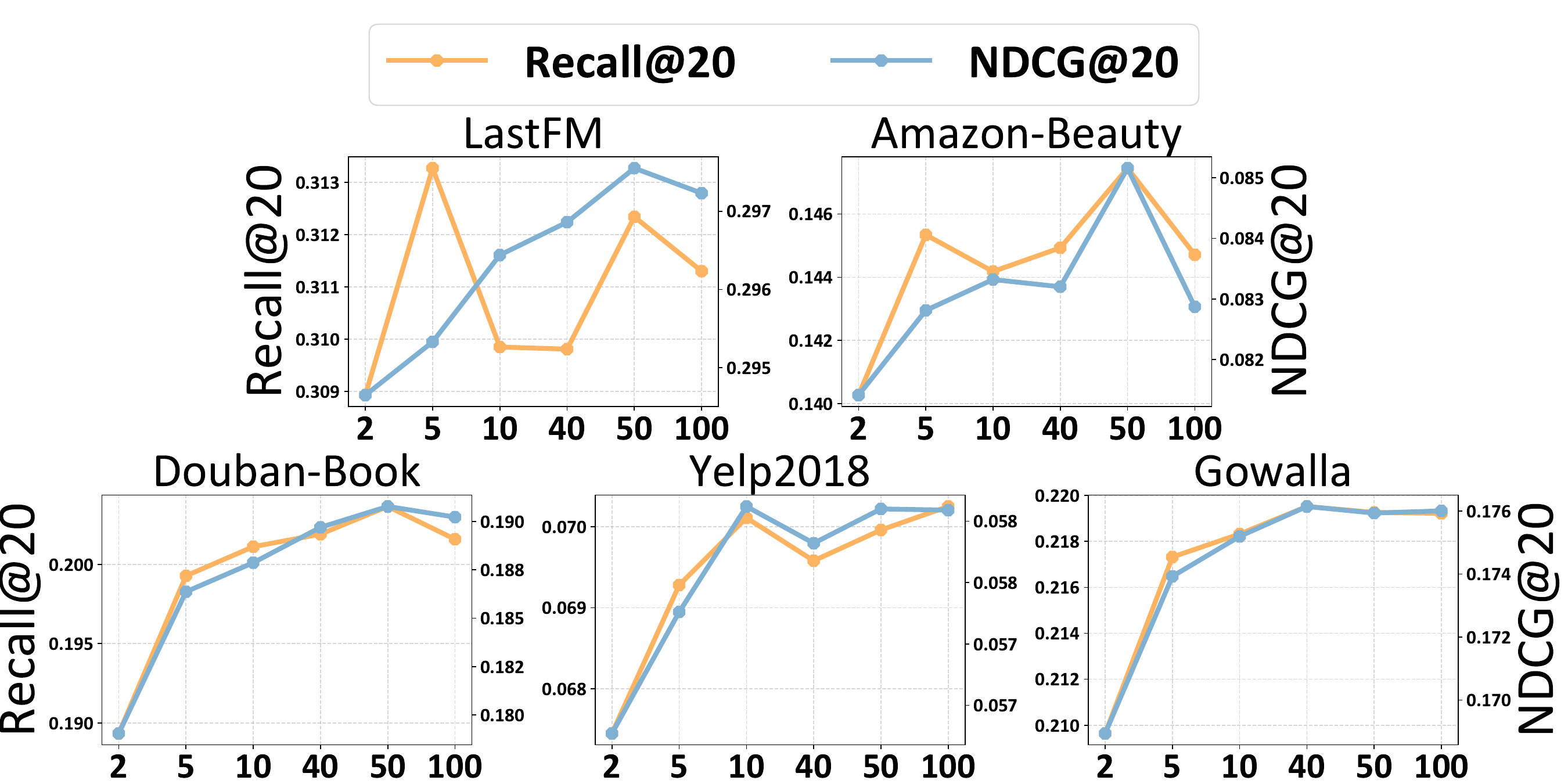}
  \caption{Influence of the number of diffusion timesteps $T$.}
  \vspace{-2mm}
  \label{fig:step}
\end{figure}

\begin{figure}[!ht]
  \centering
  \vspace{-4mm}
  \includegraphics[width=1.\columnwidth]{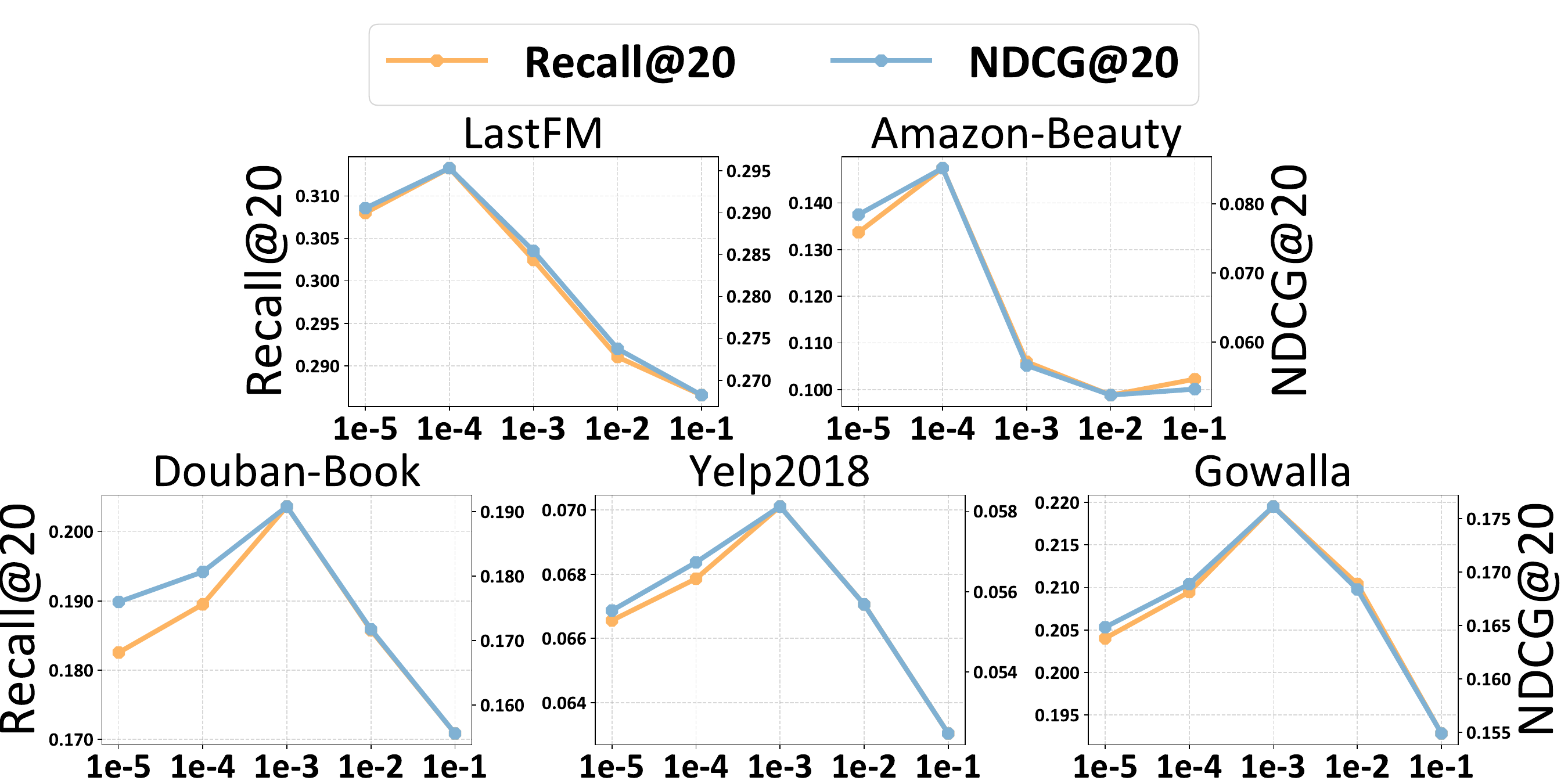}
  \caption{Influence of the noise scale $s$.}
  \vspace{-6mm}
  \label{fig:scale}
\end{figure}

\section{Supplementary Experiments}
\noindent \textbf{Hyperparameter Sensitivity Analysis for diffusion timestep $T$.}  
We analyze the sensitivity of the key hyperparameter in diffusion models: diffusion timestep $T$. This parameter bounds the maximum step for the forward and reverse processes. We vary $T$ in the range \{2, 5, 10, 40, 50 100\} as designed in~\cite{DiffRec}. According to the results (cf. Figure~\ref{fig:step}), TV-Diff consistently reaches a peak when $T=50$ for both small datasets and large datasets. This phenomenon suggests that a large diffusion timestep helps better simulate the denoising process, facilitating the underlying score-matching mechanism to capture the comprehensive distributions.

\noindent \textbf{Hyperparameter Sensitivity Analysis for noise scale $s$.}  
We also analyze the sensitivity of another key hyperparameter in diffusion models: noise scale $s$. This parameter controls the magnitude of the added noise in different timesteps. We vary $s$ in the range \{1e-5, 1e-4, 1e-3, 1e-2, 1e-1\} as designed in~\cite{DiffRec}. According to the results (cf. Figure~\ref{fig:scale}), TV-Diff reaches a peak when $s=1e-4$ and $s=1e-3$ for small datasets and large datasets, respectively. This phenomenon suggests that a small noise scale helps diffusion models to learn the nuanced distribution for small datasets, while large datasets exhibit robustness to a large noise scale.

\begin{figure}[ht]
  \centering
  \vspace{-2mm}
  \includegraphics[width=1.\columnwidth]{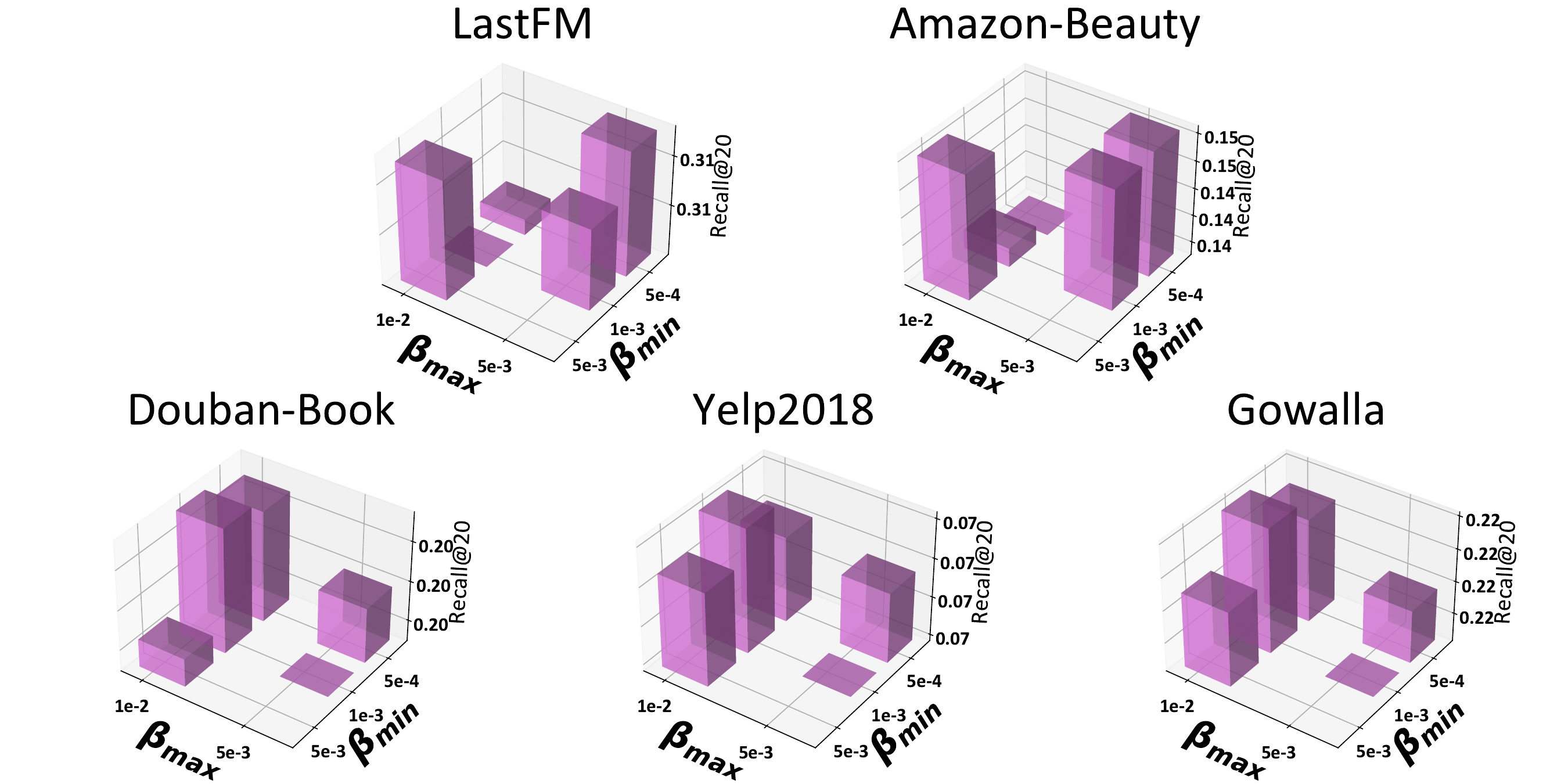}
  \vspace{-4mm}
  \caption{Influence of the upper and lower bound of noise.}
  \vspace{-2mm}
  \label{fig:min_max}
\end{figure}

\noindent \textbf{Hyperparameter Sensitivity Analysis for the upper \& low bound of noise.}  
We finally analyze the sensitivity of the last key hyperparameters in diffusion models: the upper and lower bounds of the noise. These parameters constrain the maximum and minimum noise among the whole processes. We vary them in the range of the $(\beta_{min}, \beta_{max})$ pairs \{(5e-4, 5e-3), (5e-4, 1e-3), (1e-3, 5e-3), (1e-3, 1e-2), (5e-3, 1e-2)\} as designed in~\cite{DiffRec}. According to the results (cf. Figure~\ref{fig:min_max}), TV-Diff reaches a peak when the pair $(\beta_{min}, \beta_{max})$ = (5e-4, 5e-3) and $(\beta_{min}, \beta_{max})$ = (1e-3, 1e-2) for small datasets and large datasets, respectively. This phenomenon suggests that a small difference between the upper and lower bounds helps diffusion models to maintain the consistency of learned distribution for training stability on small datasets, while a large difference provides a variety of mutual information on extreme noise to explore the credible denoising patterns on large datasets.

\begin{figure}[ht]
  \centering
  \vspace{-2mm}
  \includegraphics[width=1.\columnwidth]{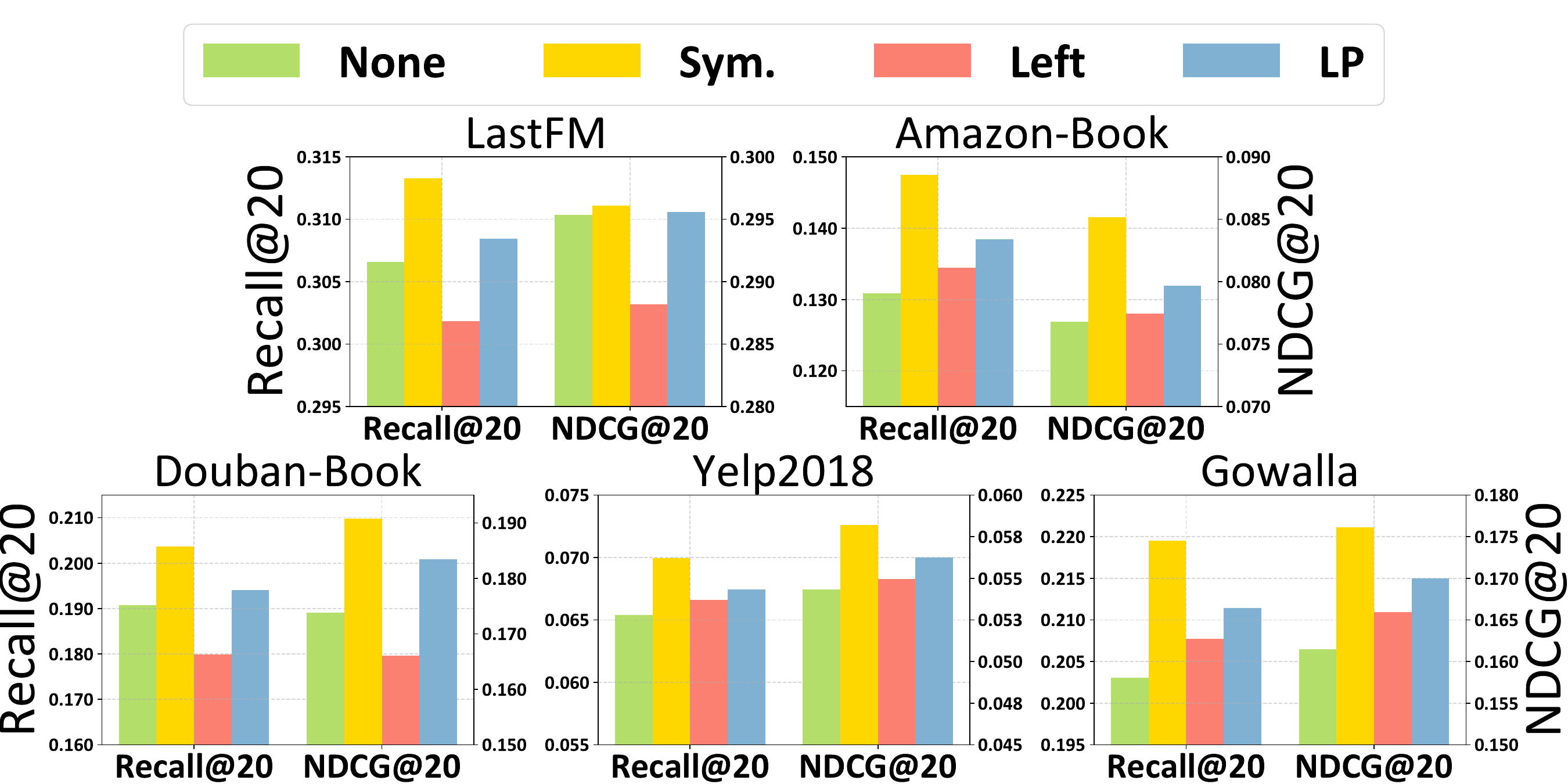}
  \caption{Influence of topological information.}
  \vspace{-2mm}
  \label{fig:graph}
\end{figure}

\noindent \textbf{Impact of Topological Information.}
We evaluate how incorporating topological information (i.e., symmetric normalization, left normalization, LinkProp~\cite{LinkProp}) affects the anisotropic denoiser. According to the results~\ref{fig:graph}, the symmetric normalized bipartite matrix consistently outperforms others, signifying its concise and accurate encoding of anisotropic user and item degree signals. In contrast, the left-normalized matrix fails to involve the item degree information, and the smoothed graph matrix of LinkProp attenuates the anisotropic signals, both resulting in model deterioration. 

\section{Algorithms}
\vspace{-1mm}
\begin{algorithm}[H] 
    \caption{The training process with TV-Diff}
    \raggedright
    \label{alg:algorithm1}
    \textbf{Input}: binary interaction $\textbf{\textit{X}}_0$, symmetric normalized $\bar{\textbf{\textit{A}}}$, \textit{entropy}-based training objective $\mathcal{L}_{S}$, temperature $t$, negative factor $\gamma$, diffusion-based hyperparameter $(T, s, \beta_{min}, \beta_{max})$.\\
    \textbf{Output}: overall learnable parameters $\theta$\\
    \begin{algorithmic}[1]
        \While{$\mbox{not convergent}$}
            \For{each batch of users $\textbf{\textit{u}}$}
                \State Sample $t\sim Uniform(1,T)$;
                \State Computer $\textbf{\textit{X}}_{t}[\textbf{\textit{u}}] \gets$Eq. (\ref{eq:q0});
                \State Reconstruct the $\textbf{\textit{X}}_{0}[\textbf{\textit{u}}]$ with $\bar{\textbf{\textit{A}}} \gets$ Eq. (\ref{eq:denoise1}), (\ref{eq:denoise2}); \textcolor{cyan}{\footnotesize\Comment{View II}}
                \State Sample negative items $j\gets$Eq. (\ref{eq:AR}), (\ref{eq:GSP});
                \textcolor{cyan}{\footnotesize\Comment{View III}}
                \State Compute $\mathcal{L}_H$ by involving $\mathcal{L}_S \gets$ Eq. (\ref{eq:helmholtz});  \textcolor{cyan}{\footnotesize\Comment{View I}}
                \State Update by descending gradients $\nabla_\theta\mathcal{L}_H$;
            \EndFor
        \EndWhile
        \State return $\theta$
    \end{algorithmic}
\end{algorithm}
\vspace{-1mm}
\begin{algorithm}[H]
    \caption{The inference process with TV-Diff}
    \raggedright
    \label{alg:algorithm2}
    \textbf{Input}: binary interaction $\textbf{\textit{X}}_0$, symmetric normalized $\bar{\textbf{\textit{A}}}$, diffusion-based hyperparameter $(T, s, \beta_{min}, \beta_{max})$, optimized parameters $\theta$\\
    \textbf{Output}: reconstructed interaction prediction $\tilde{\textbf{\textit{X}}}_0$\\
    \begin{algorithmic}[1]
        \If{$s>0$}
            \State Sample noise from $\mathcal{N}$(0, \textbf{\textit{I}});
        \EndIf
        \State Computer $\textbf{\textit{X}}_{T}$ with noise (if existing) $\gets$Eq. (\ref{eq:q0});
        \For{each timestep $t$ in $[T,\ldots, 1]$}
            \State Compute $\tilde{\textbf{\textit{X}}}_{t-1}$ by $q(\tilde{\textbf{\textit{X}}}_{t-1}|\tilde{\textbf{\textit{X}}}_{t},\textbf{\textit{X}}_{0})$ $\gets$ Eq. (\ref{eq:denoise1}),(\ref{eq:denoise2});
        \EndFor
        \State return $\tilde{\textbf{\textit{X}}}_0$
    \end{algorithmic} 
\end{algorithm}

\end{document}